\documentclass{article}

\usepackage[english]{babel}

\usepackage[a4paper,top=2cm,bottom=2cm,left=3cm,right=3cm,marginparwidth=1.75cm]{geometry}

\usepackage[mathscr]{euscript}
\usepackage{amsmath}
\usepackage{bm}
\usepackage{amsfonts}
\usepackage{bbm}
\usepackage{graphicx}
\usepackage[colorlinks=true, allcolors=blue]{hyperref}

\usepackage{authblk}

\usepackage[normalem]{ulem}
\usepackage{xcolor} 

\definecolor{dartgreen}{RGB}{0, 105, 62}

\newcommand{\add}[1]{\textcolor{blue}{#1}}

\title{Scalar embedding of temporal network trajectories}
\author{Lucas Lacasa\thanks{Corresponding author. Email: \url{lucas@ifisc.uib-csic.es.}}$^1$, F. Javier Marín-Rodríguez$^1$, Naoki Masuda$^{2,3,4}$ and Lluís Arola-Fernández}

\affil[1]{Institute for Cross-Disciplinary Physics and Complex Systems (IFISC, CSIC-UIB), 07122 Palma de Mallorca (Spain)}

\affil[2]{Department of Mathematics, State University of New York at Buffalo, Buffalo, NY, USA}

\affil[3]{Institute for Artificial Intelligence and Data Science, State University of New York at Buffalo, Buffalo, NY, USA}

\affil[4]{Center for Computational Social Science, Kobe University, Kobe, Japan}

\date{}
\begin{document}
\maketitle

\begin{abstract}
A temporal network --a collection of  
snapshots recording the evolution of a network whose links appear and disappear dynamically-- can be interpreted as a  
trajectory in graph space. In order to characterize the complex dynamics of such trajectory via the tools of time series analysis and signal processing, it is sensible to preprocess the trajectory by embedding it in a low-dimensional Euclidean space. Here we argue that, rather than the topological structure of each network snapshot, the main property of the 
trajectory that needs to be preserved in the embedding is the relative graph distance between 
snapshots. This idea  
naturally leads to dimensionality reduction approaches that explicitly consider relative distances, such as Multidimensional Scaling (MDS) or identifying the distance matrix as a feature matrix in which to perform Principal Component Analysis (PCA). This paper provides a comprehensible methodology that illustrates this approach. Its application to a suite of generative network trajectory models and empirical data certify that nontrivial dynamical properties of the network trajectories are preserved already in their scalar embeddings, what enables the possibility of performing time series analysis in temporal networks. 
\end{abstract}

\section{Introduction}
If complexity emerges out of the interactions of elements, then it is safe to say that {Network Science}  \cite{latora2017complex, newman2018networks} studies the architecture of complexity. In a nutshell, the interaction backbone of complex systems can be mathematically modeled as graphs, or more generally networks if these graphs model real-world interactions (from now on we will use the terms graph and network in an interchangeable way).  {In many occasions,} these interactions can vary dynamically, and, accordingly, networks can evolve over time. The area of temporal networks \cite{MasudaLambiotteBook, HolmeSaramaki_2012, Holme_2015, holme2019temporal} englobes such idea, and focuses on understanding how 
dynamical processes --from diffusion \cite{masuda2013temporal, delvenne2015diffusion, scholtes2014causality}, social \cite{Starnini:2013} or financial interactions \cite{mazzarisi2020dynamic} to epidemic spreading \cite{hiraoka2018correlated, van2013non},  brain activity \cite{thompson2017static} or even propagation of delays in the air transport system \cite{zanin2009dynamics, zanin2013modelling}-- are affected when the {network} changes over time \cite{williams2019effects}. While the field 
has been predominantly driven by studies of the dynamics {\it on} the network, 
more recently some focus has been paid to study, from a principled viewpoint, the intrinsic dynamics {\it of} temporal networks. The rationale is that to make use of the toolkit of dynamical systems, time series analysis, and signal processing as a means to characterize the intrinsic dynamics of a temporal network, it is helpful to interpret such temporal network as a network trajectory \cite{Correlations2022Lucas}. \\

\noindent One direction to deal with network trajectories is to develop methods that extend classical properties of time series to the network realm. In this line, classical dynamical concepts such as linear correlation functions \cite{Correlations2022Lucas, hartle2024autocorrelation, bauza2023characterization, andres2024detecting}, Lyapunov exponents \cite{Annalisa, danovski2024dynamical, caligiuri2024characterising}, or memory \cite{williams2022shape} have recently been extended to temporal networks.

\medskip \noindent 
The opposite direction, followed in this work, is to convert network trajectories into low-dimensional signals where classical methods can be {readily} applied. In this case, it is clear that simple symbolization cannot work due to high dimensionality \cite{williams2022shape}, and thus we need to resort to embedding techniques.
Graph embedding methods leverage dimensionality reduction techniques to build a projection of the nodes (or edges \cite{EdgeEmbedding}) of a single, static network in a (low-dimensional) space. {Classical} techniques include Laplacian eigenmaps \cite{belkin2001laplacian}, locally linear embedding (LLE) \cite{roweis2000nonlinear}, {and} graph factorization \cite{ahmed2013distributed} among others (see \cite{goyal2018graph} and references therein for a review). {More recently, approaches that build on network sparsity have been proposed, such as LINE \cite{tang2015line} and HOPE \cite{ou2016asymmetric}. The advent of modern deep learning has also percolated in graph embedding methods, by leveraging nonlinear dimensionality reduction such as Structural Deep Network Embedding (SDNE) \cite{wang2016structural} among others \cite{Gao_new}. From a taxonomical point of view, most network embedding methods can be categorized into three main approaches \cite{goyal2018graph}: (i) factorization-based methods, such as Laplacian Eigenmaps \cite{belkin2001laplacian} or LLE  \cite{roweis2000nonlinear}, decompose network matrices to extract latent features but struggle with modeling complex relationships. (ii) Random-walk-based methods, such as DeepWalk \cite{perozzi2014deepwalk} and node2vec \cite{grover2016node2vec}, efficiently capture local structures but require careful hypertuning. Finally (iii) Deep learning-based methods, including SDNE \cite{wang2016structural}, offer flexibility and scalability at the expense of losing interpretability and requiring large datasets and extensive training.}\\
Extensions of graph embedding ideas to temporal networks have predominantly focused again on projecting nodes \cite{Nguyen2018ContinuousTimeDN, LiNetworkEmbedding}. 
{Interestingly, only} very recently some 
approaches \cite{MasudaEmbeddingTN, Masuda_2019} have been proposed to project full network snapshots --rather than 
individual nodes or edges--.

\medskip \noindent 
Our rationale for projecting network snapshots rather than their microscopic properties is that 
each snapshot, conceived as a point in graph space, can somehow be seen as lacking internal structure \cite{Correlations2022Lucas}. 
Moreover, if one aims to do time series analysis --or signal processing-- of network trajectories, then the key aspect of the network trajectory to be preserved in the embedding is the relation between snapshots, rather than the relation between the nodes or edges of each snapshot. {Note that a similar insight, incidentally, is at the core of the well-known visibility graph by which information stored in time series can be efficiently mapped into a graph-theoretical representation \cite{lacasa2008time}.}
Such insight further suggests considering (quasi)-isometric transformations of the temporal network, which recently led to the proposal \cite{MasudaEmbeddingTN} that explores the performance of {m}ultidimensional {s}caling (MDS) embedding of a specific model of temporal networks called {tie-decay networks} \cite{MasudaEmbeddingTN, thongprayoon2023online, thongprayoon2024spline}. {However,} these  works do not explicitly address whether low-dimensional embedding preserve{s} key statistical properties of network trajectories, and to which extent standard concepts like memory, temporal correlations or dynamical instability 
can be retrieved from such low-dimensional embeddings.\\

\noindent Here, we expand on \cite{MasudaEmbeddingTN} {to} consider various possible approaches one can follow to obtain low-dimensional --and in particular, scalar-- embeddings of network trajectories using linear dimensionality reduction methods. We build various types of complex dynamics, from one-dimensional processes to synthetic network trajectories --white, noisy periodic, autorregressive, chaotic-- and mixtures thereof, and complement these with empirical temporal networks. We systematically explore how correctly the resulting embeddings capture the subtle, intrinsic dynamics of the original network trajectory. {To that aim, we use graph metrics that characterise specific dynamical properties of network trajectories, such as periodicity and memory of sensitivity to initial conditions.}\\
The rest of the paper goes as follows. In Sec. \ref{sec:methods}, we define our methodology, which encompasses four strategies to obtain low-dimensional (and in particular, scalar) embeddings of network trajectories that leverage two different dimensionality reduction philosophies: principal component analysis and multidimensional scaling \cite{cox2000multidimensional}. In this section, we also define the metrics used to validate our results, which include autocorrelation functions, Lyapunov exponents, and their extensions for network trajectories.
In Sec. \ref{sec:results}, we describe the results of applying the embedding methodology to signals of different complexity, ranging from 1D processes to synthetic temporal network models. We also apply the method to a couple of empirical temporal networks, in order to showcase how the method works in real scenarios. In Sec. \ref{sec:discussion} we conclude and discuss open problems for future work.

\section{Methodology}
\label{sec:methods}

Let us define a temporal network --or network trajectory-- as an ordered sequence of $T$ graphs ${\cal S} = (G_1,G_2,\dots,G_T)$, where $G_t$ is the $t$-th network {\it snapshot}. When the nodes are labeled, $G_t$ can be fully represented by its adjacency matrix ${\bf A}(t)$, with entries $A_{ij}(t)$. We assume that each network snapshot has a fixed number of nodes $N$ and let the time-evolving interactions be weighted or directed in general. 
We define the low-dimensional Euclidean embedding of such trajectory as a time series ${\cal S}_{\Phi}:=\{z_1,z_2,\dots,z_T\}$, where $z_t \in \mathbb{R}^\textsc{dim}$, and, in general, we aim at $\textsc{dim}\ll T$ ($\textsc{dim}=1$ is the case of special interest that produces a scalar embedding). Accordingly, the embedding function $\Phi: G_t\to z_t$ assigns a point $z\in \mathbb{R}^\textsc{dim}$ to every graph: $z = \Phi(G), \  \forall \ G \in {\cal S}$. The whole problem, therefore, reduces to find the function $\Phi(\cdot)$. \\

\noindent In this work, we construct functions $\Phi(\cdot)$ based on linear dimensionality reduction schemes. In particular, we consider four strategies to build $\Phi(\cdot)$ based either on PCA or MDS. The proposed methods are conceptually similar {to each other}, since all apply dimensionality reduction to the network trajectory, but differ in a few technical details which produce some variability in the results and highlight different aspects of the problem.

\subsection{PCA-based strategies} 
\label{sec:PCA}
Principal Component Analysis (PCA)\footnote{PCA, or slight variations of it, receives other names depending on the specific field of application, e.g. Proper Orthogonal Decomposition (POD) or Factor Analysis, among others.} is a widely used linear dimensionality reduction technique that identifies orthogonal directions along which the variance of the data is maximized \cite{jolliffe2002pca}. Formally, PCA involves the spectral decomposition of a covariance matrix of data features in terms of the eigenvalues (which capture the magnitude of the explained variance) and the associated eigenvectors (the principal components). \\ 


\noindent In order to apply PCA to our problem, the network snapshots $G_t$ are originally projected in a suitable (Euclidean) feature space. Now, {how shall we define the feature vector of each network snapshot? Shall we just extract a list of network scalar metrics associated to each snapshot? Shall we focus on edge-based metrics? Node-based ones? Shall we use all the entries of the adjacency matrix as features? A key insight is that, when we aim at preserving dynamical properties of the whole network {\it trajectory}, then it is sensible to focus on the {\it relative position} of each network snapshot in a graph space, rather than considering specific topological information about each network snapshot.} In other words, 
an intuitive solution is to 
consider the set of relative distances between a snapshot $G_t$ and every other snapshot as the {\it features} of snapshot $G_t$. Accordingly, $G_t$ is expressed as a vector of $T$ features, where the $j$-th feature depicts the distance between $G_t$ and the $j$-th network snapshot. Subsequently, one can project $G_t$ in such a feature space, where the first axis relates to the distance of a generic snapshot $G_t$ to $G_1$, the second axis to the distance to $G_2$, and so forth. By taking the spectral decomposition of the distance-based covariance matrix and projecting the data points into the first principal component (or by simply using the first component), we obtain a scalar embedding of the network trajectory. \\

\noindent More specifically, we first define the features of each network snapshot based on pairwise squared distances. From pairs of snapshots \( G_t \) and \( G_\ell \), we construct the squared distance matrix:
\begin{equation}
    \mathcal{D}^{(2)} = \{d_{t\ell}^2\}_{t,\ell=1}^T, \quad d_{t\ell}^2 = \|G_t - G_\ell\|^2,
\end{equation}
where \( \|\cdot\| \) is a suitable norm (e.g., Frobenius or an $L_p$ norm for adjacency matrices). While using standard distances $d_{ij}$ already produce{s} decent results in our problem, squared distances\footnote{which correspond to use as feature matrix ${\cal D}^{(2)}= {\cal D} \odot {\cal D}$, where $\odot$ is the Hadamard, entrywise product.} are a better choice because they have a direct relationship with the inner product space and preserve better the distances after the spectral decomposition is applied (as clarified in the MDS section). Before applying such decomposition, in PCA the covariance matrix must be column-centered, such that features have zero mean and variance can be maximized along the principal components \cite{jolliffe2002pca}:
\begin{equation}
    \tilde{\mathcal{D}}_{t\ell}^{(2)} = d_{t\ell}^2 - \langle d_{t\ell}^2 \rangle_\ell,
\end{equation}
where \( \langle d_{t\ell}^2 \rangle_\ell \) is the mean over columns. 
The column-centered matrix \(\tilde{\mathcal{D}}^{(2)}\) serves as the input feature space for PCA. Since the covariance matrix $\tilde{{\cal D}}^{{(2)}^\top} \tilde{{\cal D}}^{(2)}$ is always symmetric and positive definite, the spectral decomposition:
\begin{equation}
\tilde{{\cal D}}^{(2)^\top} \tilde{{\cal D}}^{(2)} = \sum_{i=1}^T \Lambda_i {\bf e}_i{\bf e}_i^\top
\end{equation}
has always non-negative eigenvalues which can be ordered as $\Lambda_1\ge \Lambda_2\ge\dots\ge\Lambda_T\ge 0$ and $T$ associated orthogonal and real eigenvectors, where ${\bf e}_i=(e^1_i,e^2_i,\dots,e^T_i)$ is the $i$-th eigenvector --also called the $i$-th principal component-- with entries $e_i^j \in \mathbb{R}$. Reducing the dimensionality of each network snapshot $G_t$ from $T$ (the original dimension of the feature set) to $\textsc{dim}\ll T$ implies 
truncating this decomposition at order $\textsc{dim}$. From this point, we identify two slightly different strategies of finding a low-dimensional embedding of the network trajectory $\cal S$:\\

\noindent {\it i) PCA-projection strategy}: The most conventional approach in PCA is to systematically project the feature vector of each network snapshot onto the first $\textsc{dim}$ principal components. If we label ${\bf v}_t$ as the feature vector of $G_t$ (the $t$-th row in $\tilde{{\cal D}}^{(2)}$), then $z_t=({\bf v}_t \cdot {\bf e}_1, {\bf v}_t \cdot {\bf e}_2,\dots, {\bf v}_t \cdot {\bf e}_{\textsc{dim}})$, or in matrix form $\mathbf{Z} = \tilde{{\cal D}}^{(2)}\mathbf{E}_{\textsc{dim}}$, where $\mathbf{E}_{\textsc{dim}} \in \mathbb{R}^{T \times \textsc{dim}}$ is a matrix containing the first $\textsc{dim}$ eigenvectors as columns. In the particular case of a scalar embedding ($\textsc{dim} = 1$), we get: 
\begin{equation}
z_t={\bf v}_t \cdot {\bf e}_1,
\end{equation}
meaning that the scalar embedding of snapshot $G_t$ is just the inner product of the {feature vector} of that snapshot and the first eigenvector of the decomposition. \\

\noindent {\it ii) PCA-embedding strategy}: An alternative and non-standard approach is to directly identify the entries of the scaled principal components (weighted by the square root of the eigenvalues), with the embedded coordinates. Anecdotically, this is similar in spirit {to} some methods in graph-based spectral embedding. 
In this strategy, the $\textsc{dim}$-order embedding of $G_t$ corresponds to the $t$-th component of the set of $\textsc{dim}$ (scaled) eigenvectors $z_t=(\sqrt{\Lambda_1}{e}_1^t, \sqrt{\Lambda_2}{e}_2^t, \dots, \sqrt{\Lambda_\textsc{dim}}{e}_{\textsc{dim}}^t) \in \mathbb{R}^{\textsc{dim}}$, where ${e}_k^t$ stands for the $t$-th entry of the vector ${\bf e}_k$, or in matrix form $\mathbf{Z} = \mathbf{E}_{\textsc{dim}} \mathbf{\Lambda}_{\textsc{dim}}^{1/2}$ where $\mathbf{\Lambda}_{\text{dim}}^{1/2} = \text{diag}(\sqrt{\Lambda_1}, \sqrt{\Lambda_2}, \dots, \sqrt{\Lambda_{\textsc{dim}}})$. In the scalar case, the embedding of $G_t$ simplifes to: 

\begin{equation}
z_t = \sqrt{\Lambda_1}e_1^t, 
\end{equation}
where all the information required for the embedding is contained in the first (scaled) eigenvector.\\ 

\noindent Both strategies rely on the same decomposition but differ in interpretation. While the  \textit{PCA-projection strategy} emphasizes variance in the feature space (where distances are squared), the \textit{PCA-embedding strategy} directly leverages the spectral decomposition, potentially better preserving pairwise relationships and aligning more closely with the geometry of the network snapshots.

\subsection{MDS-based strategies}
\label{sec:MDS}
An a priori more direct approach is to make use of a spectral truncation method that, by construction, aims to preserve as much as possible the pairwise distance between points: we aim at building a quasi-isometrical transformation that reduces the dimensionality. This is the remit of the family of {m}ultidimensional {s}caling (MDS) algorithms \cite{borg2005mds, cox2000multidimensional}, 
{used in} {previous} work on tie-decay network embedding \cite{MasudaEmbeddingTN}. As in the PCA case, we consider two different strategies {based on MDS}, both of them being conceptually similar and aiming at the same goal but differing  in  technical {details}.\\

\noindent {\it iii) Classical-MDS strategy}: The classical idea of MDS is to reconstruct a hidden inner product space from the squared distances between points \cite{borg2005mds}. In fact, the connection between squared distances and inner products is key to understanding how MDS works and its relationship to the previous PCA-based strategies. If  $G_t$ and $G_\ell$ {are} two generic snapshots, with pairwise squared distance $d_{t\ell}^2$, this squared distance can be formally expressed as the inner product space of some latent (hidden, i.e., unobservable) features as:
\begin{equation}
    d_{t\ell}^2 = \|\mathbf{x}_t - \mathbf{x}_\ell\|^2 = \|\mathbf{x}_t\|^2 + \|\mathbf{x}_\ell\|^2 - 2 \mathbf{x}_t \cdot \mathbf{x}_\ell,
\end{equation}
where $\mathbf{x}_t \in \mathbb{R}^Q$ is the unknown feature vector of snapshot $G_t$ (in our case, if the feature vector $\mathbf{x}_t$ were to correspond to the full adjancency matrix $\mathbf{A}(t)$, then $Q = N \times N$.), and $\|\mathbf{x}_t\|^2$ is its squared norm. This formal relationship indeed allows reconstructing the inner product matrix $\mathcal{B}$, which encodes the geometry of the hidden feature space. The trick works by first applying a double-centering transformation to the squared-distance matrix $\mathcal{D}^{(2)}$, which includes the column-centering of the PCA approach and also a row-centering and finally a global-centering across the whole matrix. This transformation can be compactly written in matrix form as $\mathbf{J} = \mathbf{I} - \frac{1}{T} \mathbf{1} \mathbf{1}^\top$\add{,}
where $\mathbf{I}$ is the identity matrix and $\mathbf{1}$ is a column vector of ones. The double-centered matrix $\mathcal{B}$ is given by:
\begin{equation}
    \mathcal{B} = -\frac{1}{2} \mathbf{J} \mathcal{D}^{(2)} \mathbf{J}.
    \label{B_matrix}
\end{equation}
Remarkably, this step reconstructs the inner product matrix exactly since $\mathcal{B} = \mathbf{X} \mathbf{X}^\top \in \mathbb{R}^{T \times T}$, where $\mathbf{X}$ contains the unknown feature vectors $\mathbf{x}_t$ as rows. The matrix $\mathcal{B}$ captures the geometry of the (hidden) feature space only using a properly centered square distance matrix, instead of using the full information of the (unknown) feature vectors. Finally, to get the low-dimensional embeddings, we truncate the spectral decomposition of ${\cal B}=\sum_{k=1}^T\lambda_k {\bf u}_i{\bf u}_i^\top$ up to $\textsc{dim} \ll T$. The embedding of each snapshot $G_t$ is directly given by the entries of the first $\textsc{dim}$ eigenvectors of $\mathcal{B}$, scaled by the square root of the corresponding eigenvalues. In the scalar case, the embedding of $G_t$ is 
\begin{equation}
    z_t = \sqrt{\lambda_1} u_{1}^t.
\end{equation}
Notice the resemblance between this approach and performing PCA on the squared-distance matrix. In fact, this strategy is 
{algebraically equivalent} to the {\it PCA-embedding strategy}, the only difference being that the column-centering of the squared-distance matrix in PCA becomes a double-centering in MDS. Since the latter is the transformation that better preserves the distances (instead of the variances), one might expect that this method should consistently outperform PCA-based approaches, although this is not systematically the case (see e.g. Fig.\add{~}\ref{fig:DARN3ACF} for a case where the properties of network trajectories with planted short-term memory are better captured by the PCA embedding than the {c}lassical MDS one). 
\\

\noindent {\it iv) Metric-MDS strategy}: Finally, an alternative approach is to make use of the so-called metric MDS method, an embedding which explicitly aims at minimizing the mismatch between the pairwise distances in the embedded space and the original distance matrix. This involves solving an optimization problem to minimize the so-called stress function:
\begin{equation}
    \text{Stress} = \sum_{t,\ell=1}^T \left( d_{t\ell} - \|z_t - z_\ell\| \right)^2,
\end{equation}
where $z_t$ and $z_\ell$ are the embedded coordinates of snapshots $G_t$ and $G_\ell$, respectively, and $d_{t\ell}$ is their original distance. Unlike classical MDS, this method does not explicitly rely on a spectral decomposition but instead uses iterative optimization techniques to find the embedding that minimizes the stress \cite{borg2005mds}. 
{As we will show, this strategy introduces spurious effects and is thus less efficient than the other three.}\\

\subsection{Validation strategies}


In order to validate the working hypothesis {that preserving relative distances between snapshots enables to build accurate scalar embeddings via the methods proposed above}, in this work we have studied the performance of the four methodological strategies discussed {in Secs.~\ref{sec:PCA} and \ref{sec:MDS}} in building scalar embeddings of (i) complex one-dimensional dynamics, and (ii) temporal network trajectories of varied complexity. {We use}  three tools {to validate the methods}: When the original trajectory has a canonical embedding (e.g. when the original trajectory is itself a {one-dimensional} time series, and thus $G_t=x_t \in \mathbb{R}$, see Sec. \ref{sec:prelim_1d}), validation is performed by assessing the Pearson correlation $r$ and the Spearman correlation $\rho$ of the scatter plot $z_t$ vs $x_t$ (correct embeddings $z_t\propto x_t$ give $r=\rho=1$).
Network trajectories on the other hand are inherently difficult to project in a low-dimensional space and thus we lack {a} direct {ground truth} to compare our embeddings. 
In this case, we resort to their statistical properties with the aid of the network extensions of autocorrelation \cite{Correlations2022Lucas} and Lyapunov exponents \cite{Annalisa} as follows:\\ 

\noindent For models of temporal networks with planted memory, we compare the network autocorrelation function $\text{nACF}(\tau)$ of the original network trajectory \cite{Correlations2022Lucas}, computed from the network trajectory as
\begin{equation}
    \text{nACF}(\tau) = \texttt{tr}\bigg(\frac{1}{T-\tau}\sum_{t=1}^{T-\tau} [{\bf A}(t)-\mu]\cdot[{\bf A}(t+\tau)^\top -\mu^\top]\bigg),
    \label{eq:nACF}
\end{equation}
where ${\bf A}(t)$ is the adjacency matrix of the $t$-th network snapshot, $\mu=\frac{1}{T}\sum_{t=1}^T {\bf A}(t)$ is the annealed {(i.e., time-averaged)} adjacency matrix of the whole temporal network trajectory, $\top$ denotes matrix transposition and $\texttt{tr}(\cdot)$ is the trace operator. $\text{nACF}(\tau)$ will play the role of our ground truth against which we will compare the autocorrelation function $\text{ACF}(\tau)$ of the scalar embedding obtained via PCA or MDS:
\begin{equation}
    \text{ACF}(\tau) = \frac{1}{\sigma_z^2(T-\tau)}\sum_{t=1}^{T-\tau} (z_t - \mu_z)(z_{t+\tau} - \mu_z),
    \label{eq:ACF}
\end{equation}
where $\mu_z = \langle z_t\rangle$ is the mean of the time series {$\{ z_1, \ldots, z_T \}$} and $\sigma^2_z = \langle z_t^2\rangle - \langle z_t\rangle^2$ is the variance. Notice that $\text{ACF}(\tau)$ is normalised between {$-1$ and $1$} but $\text{nACF}(\tau)$ is only centered, so matching should not be identical accordingly.\\

\noindent For models of temporal networks showing sensitive dependence of initial conditions, the network Maximum Lyapunov Exponent (nMLE \cite{Annalisa}) will serve as ground truth, against which to compare the estimated maximum Lyapunov exponent (MLE) of the scalar embedding. Such MLE will be found using Wolf's method, that exploits recurrences in the scalar embedding trajectory $z_t$ to build pairs of initial{ly} close surrogate trajectories whose expansion over time is estimated. Concretely, pairs of points $z_u$ and $z_v$ in the scalar embedding which are close in that space, i.e., $|z_u-z_v|<\epsilon$, are tracked. The ansatz in Wolf's method assumes that the distance function $d(k) = |z_{u+k}-z_{v+k}|$ displays exponential growth $d(k) \sim \exp(\lambda k)$, where $\lambda$ is a local expansion rate which in general depends on the position of $z_u$ (or $z_v$, depending which one is considered a perturbed condition). The MLE is simply an average of $\lambda$ over different initial conditions. 

\section{Results}
\label{sec:results}


\begin{figure}[htb!]
\centering
\includegraphics[width=0.95\linewidth]{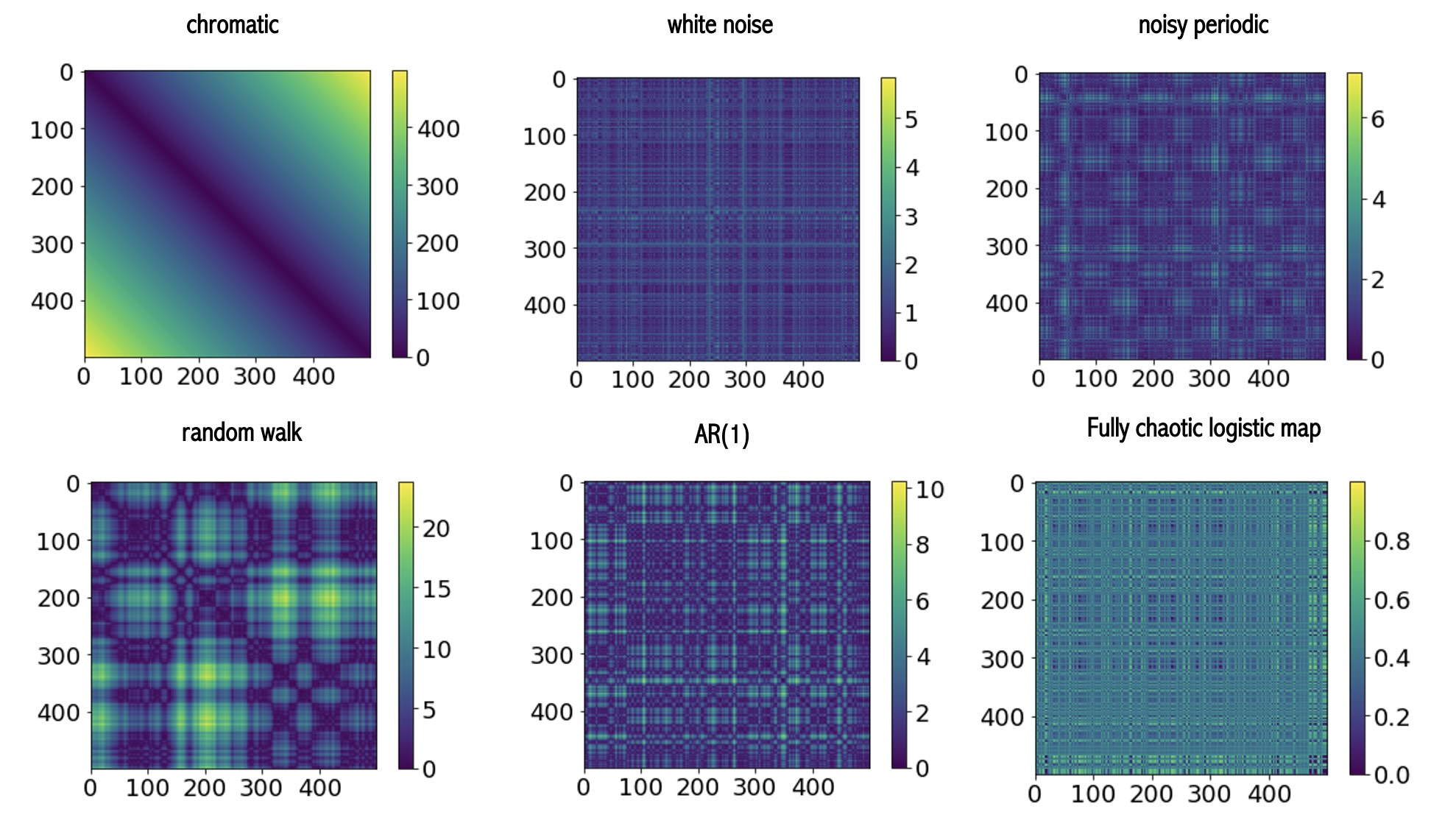}
\caption{\label{fig:1D} {\bf {One-dimensional} {v}alidation set.} Distance matrices from {one-dimensional} trajectories of different complexity, used to provide a preliminary validation of the methods.}
\end{figure}

\subsection{Preliminary validation of complex scalar time series}
\label{sec:prelim_1d}
To validate the methods, we generate time series $\{x_t\}_{t=1}^T$ from a variety of synthetic, one-dimensional processes with different {levels of complexity}.
In each case, we extract the distance matrix ${\cal D}=\{d_{ij}\}; \ d_{ij}=\vert x_i - x_j \vert$ and subsequently build the scalar embedding procedures that yield the different functions $\Phi(\cdot)$ so that the embedded scalar signal is $(z_t)_{t=1}^T$, where $z_t = \Phi (x_t)$. Since the original signal is actually one-dimensional, we assess the validity of the framework by scatter plotting $z_t$ vs $x_t$ and computing two different correlation coefficients: Pearson's and Spearman's correlation coefficients (the former captures linear correlations, while the latter captures monotonic relationships which are not necessarily linear). 
The list of synthetic processes include: (i) a chromatic process $x_t=t$, (ii) white Gaussian noise\footnote{The characteristic of this process is that its autocorrelation function is a Dirac delta centered at lag $\tau=0$.}  $x_t=\xi$, $\xi \sim N(0,1)$, (iii) a nois{y} periodic signal $x_t = \cos(2\pi t/P) + \xi$ of period $P=100$, where $\xi \sim \mathcal{N}(0,\sigma)$ is an extrinsic noise polluting the periodic signal\footnote{The autocorrelation function peaks at $\tau=P$ (and subsequent harmonics), and the signal to noise ratio affects the height of such peak.}, (iv) a Gaussian {r}andom {w}alk\footnote{This is a non-stationary process whose power spectrum is a power law.} $x_{t+1}=x_t+\xi$, $\xi \sim \mathcal{N}(0,1)$, (v) an autorreggressive process\footnote{with exponentially decaying autocorrelation function.} of order 1 $x_{t+1}=0.7x_t+\xi$, $\xi \sim \mathcal{N}(0,1)$, and (vi) a chaotic process generated by a fully chaotic logistic map\footnote{This is a deterministic 1D chaotic process whose autocorrelation function has again a Delta-shape (like white noise) but that it shows sensitivity to initial conditions: two initially close trajectories diverge exponentially fast, with a characteristic Lyapunov exponent $\Lambda = \ln (2)$.} $x_{t+1}=4x_t(1-x_t)$. In every case we generate a time series of $T=500$ points from these processes. The distance matrices are reported in Fig.~\ref{fig:1D} for illustration. 

\medskip
\noindent 
After building the different scalar embeddings, we have observed that the method based on PCA embedding is systematically successful, as shown by a perfect matching (Pearson correlation coefficient $=1$) between the original signal $\{x_t\}$ and the embedded signal $\{z_t\}$ for all {the six time series}, {implying a perfectly linear relationship} $z_t=\Phi(x_t)= \alpha x_t$, for some $\alpha \ne 0$. Anecdotically, the explained variance associated to the first principal component hovers around $90\%$ for all six time series. 
 {The perfect matching also holds for the method based on Classical-MDS. In this case, one can prove using basic linear algebraic arguments that the scalar embedding recovers the original one-dimensional signal up to a constant shift and a reflection, as shown in Appendix A.} The method based on using a PCA projection is also successful {at reproducing a monotonous relationship between the original signal and the scalar embedding}, albeit the scatter plot between $\{x_t\}$ and $\{z_t\}$ is not a straight line, but rather has a curved shape (with small curvature, so the Pearson coefficient $<1$ but it is very close to 1) and the Spearman coefficient $=1$. Finally, the method based on using metric-MDS suffers from the problem of observing the onset of what we call  ``antiphases'' in the signal: time points $t$ where, instead of having the correct assignment $z_t=\alpha x_t$, the embedding provides a sign flipping $z_t=-\alpha x_t$ (see Appendix B Fig.~\ref{fig:antiphase} for an illustration). The {frequency of these undesired antiphases} is usually between $1\%$ and $15\%$. These antiphases are clearly observable when the signals are smooth, but it is less evident otherwise. Additionally, these antiphases can break the temporal structure of the embedding. Removing these artifacts is relatively easy in the most simple case where when there is an available ground truth (see Appendix B). In general, the best one can do is to assume that the embedding is smooth and flip the sign of the points that violate such assumption, but such smoothness assumption does not necessarily hold.\\
All in all, these findings already makes the strategy based on metric-MDS less useful than the other three. {Accordingly, in what follows we focus on the other three types of embedding.}

\subsection{Synthetic network trajectories}
We now proceed to analyze synthetic network trajectories of different garment. For simplicity, we make use of the Euclidean norm and, accordingly, the distance $d({\bf A},{\bf A}')$ between two network snapshots with adjacency matrices ${\bf A}=\{A_{ij}\}$ and ${\bf A}'=\{A'_{ij}\}$ is defined as
\begin{equation}
    d({\bf A},{\bf A}') = \sqrt{\frac{1}{N^2}\sum_{i,j=1}^{N} (A_{ij}-A'_{ij})^2}.
    \label{eq:d_snapshots}
\end{equation}
{Note that the entries of the matrices do not necessarily need to be binary, hence network snapshots can be weighted and weights can be positive or negative.}


\begin{figure}[htb!]
\centering
\includegraphics[width=0.99\linewidth]{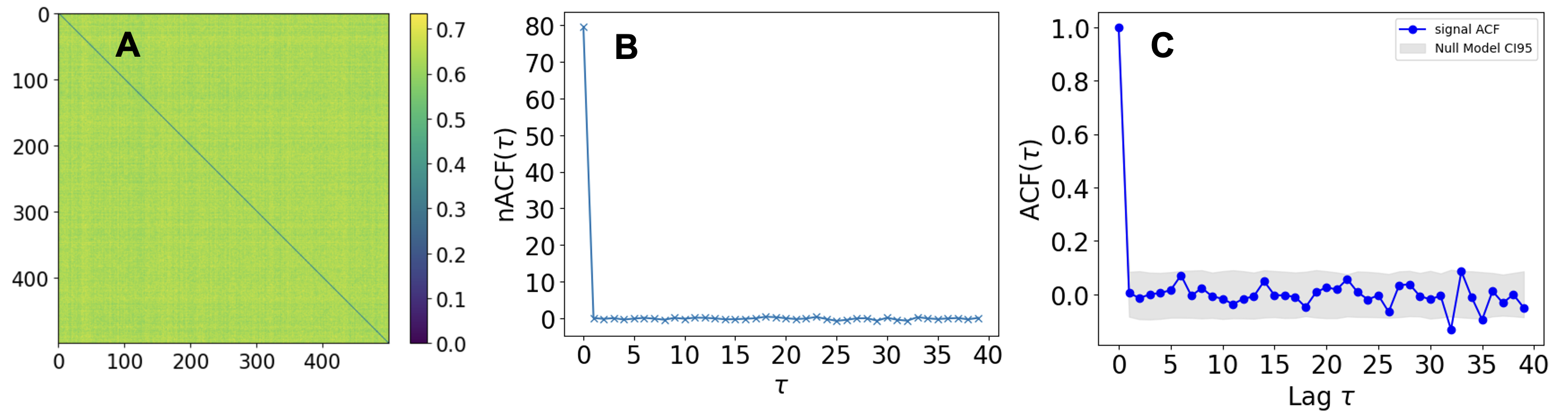}
\caption{\label{fig:white} {\bf Autocorrelation analysis of white network trajectories and its embeddings.}  (A) Distance matrix of a network trajectory of $T=500$ snapshots, where each snapshot is an Erd{\H{o}}s-R\'enyi graph of $N=20$ nodes sampled from $\mathscr{G}(20,0.3)$. (B) Network Autocorrelation Function $\text{nACF}(\tau)$ estimated from the network trajectory{,} showing the characteristic Dirac-delta shape of uncorrelated signals. 
(C) Scalar autocorrelation function $\text{ACF}(\tau)$ (alongside a $95\%$ confidence interval of a null model where the {one-dimensional}  signal was shuffled $10^3$ times) associated to the PCA-embedding, correctly capturing the uncorrelated nature of the network trajectory.  All four embeddings display a similar $\text{ACF}(\tau)$ as the one displayed in this panel.
}
\end{figure}

\subsubsection{White network trajectories}\label{sec:white}
We start by the simplest network trajectory: one emulating an uncorrelated stochastic process in graph space with a Dirac delta-shaped autocorrelation function. To do that \cite{Correlations2022Lucas}, the network trajectory $\cal S$ is built by sampling a total of $T$ Erd{\H{o}}s-R\'enyi graphs from $\mathscr{G}(N,p)$. For illustration, in Fig.~\ref{fig:white}{(A)} we show the resulting distance matrix when $T=500$, $N=20$ and $p=0.3$. {Figure~\ref{fig:white}}(B) reports the network autocorrelation function, that adequately captures the expected shape (note that $\text{nACF}(0)\ne 1$ as it is not normalized). The scalar embeddings produced by all four strategies yield the correct scalar autocorrelation function (see {Fig.~\ref{fig:white}(}C{)}). Incidentally, observe that in this uncorrelated case the onset of antiphases in the metric-MDS case do not break down the statistical properties of the signal --as the signal remains uncorrelated--, but precisely because of the non-smooth nature of this signal, applying an antiphase correction (which is based on assuming smoothness in the embedding, see Appendix B for details) introduces spurious correlations in the metric-MDS embedding. 

\medskip \noindent {At this point, we also wish to investigate whether, and to what extent, stacking additional snapshots in a network trajectory can yield a change in the scalar embedding. By construction, each of the points $z_t$ depend on all the network snapshots $G_0,G_1,\dots,G_{T-1}$, hence in principle just adding an extra snapshot $G_{T}$ can modify all scalar embeddings $z_0,z_1,\dots,z_{T-1}$. Nonetheless, we remind that the specific value of each $z_t$ is often not what matters, but rather, how these values change relative to each other. Similarly to the case of time series, interesting properties of a time series are usually invariant under translations of the whole time series.
To investigate the effects of stacking additional snapshots, starting from the white network trajectory described above, we add $k$ snapshots to build a network trajectory ${\cal S}$ composed of $T+k$ snapshots, and subsequently build the whole scalar embedding $z_t$. We plot the initial scalar $z_0=\Phi(G_0)$ as a function of $k$ in Fig.~\ref{fig:stability} in Appendix C. Results indicate that all three viable scalar embeddings (both PCA-based embeddings and classic-MDS) are stable against addition of snapshots, with the exception of the classic-MDS one that can suffer sign shifts.}

\subsubsection{Pulsating network trajectories: the scalar embedding as a noise filter}
As a second step, we consider a {weighted} temporal network model where {the weight of} each link in the network independently {and asynchronously evolves according to the following dynamics: $A_{ij}(t)=\cos(2\pi t/P + \eta_{ij}) + \xi,$ where $P$ is the period of the periodic part, $\eta_{ij}\sim \textsc{Uniform}(0,1)$ is a quenched uniform random variable that introduces asynchronous behavior in the link activities, and $\xi \sim \mathscr{N}(0,\sigma)$ is an extrinsic dynamic noise polluting each link in the network in an uncorrelated way.}
The periodic component of the individual link dynamics makes the network to collectively `pulsate'.
Observe that the standard deviation {of the dynamic noise,} $\sigma${,} modulates the signal{-}to{-}noise ratio (SNR), which goes to zero as $\sigma$ increases as $1/\sigma^2$. Here, we consider two cases: $\sigma=1$ (where the characterization of periodicity at the link level is still possible although the detection is not fully trivial) and $\sigma=4$ (where there is virtually no trace of the periodic component at the link level, and the link dynamics appears as white noise).\\

\begin{figure}[htb!]
\centering
\includegraphics[width=0.95\linewidth]{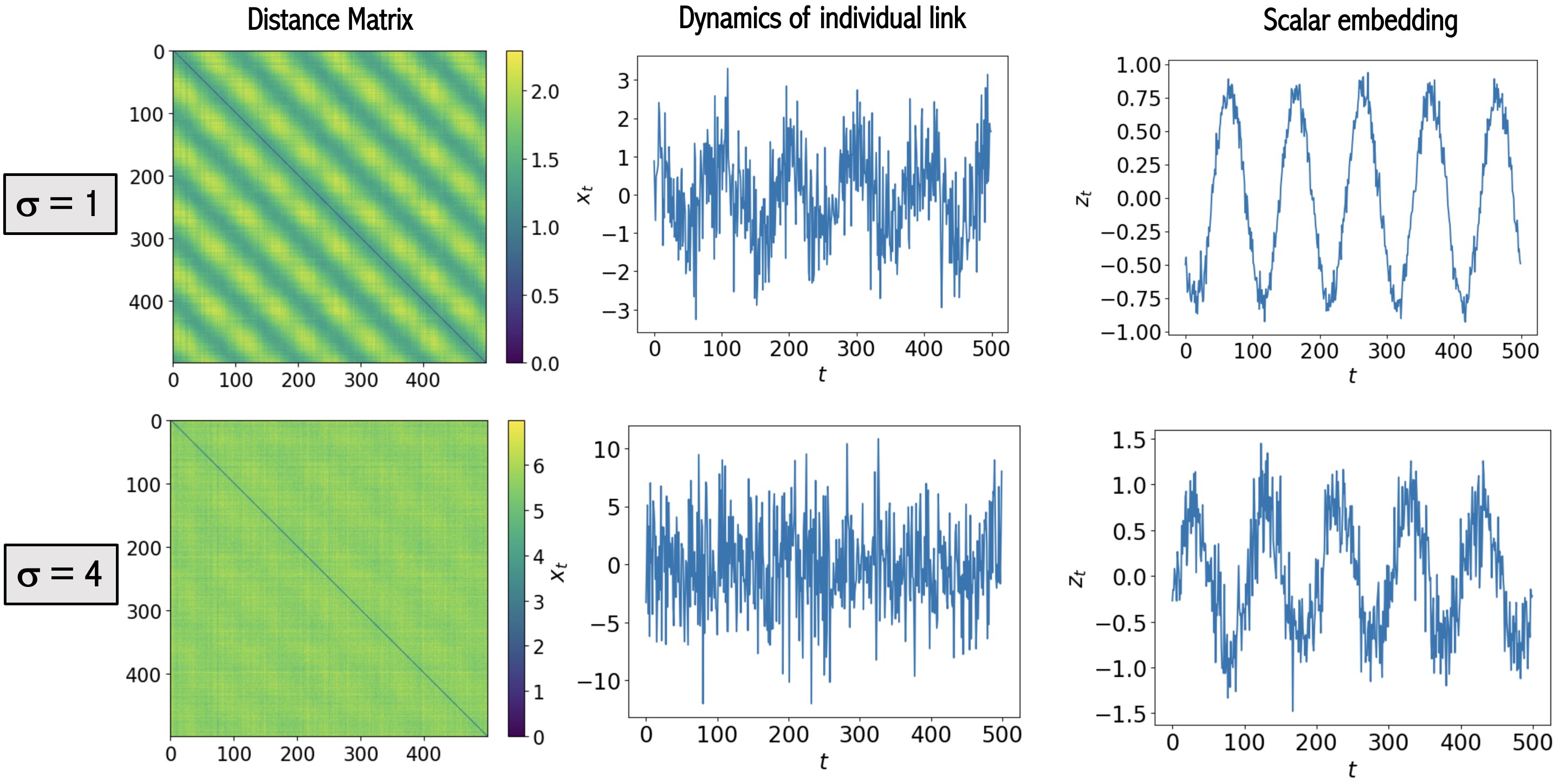}
\caption{\label{fig:pulsating} {\bf Noisy {p}ulsating network trajectories. } We generate a noisy pulsating temporal network trajectory of $N=500$ snapshots, where each link consist{s} in an asynchronous sinusoidal dynamics of period $P=100$ polluted with extrinsic Gaussian noise of standard deviation $\sigma$. The left panels display the distance matrix between network snapshots, for two cases: $\sigma=1$ and $\sigma=4$. The middle panels display the time series of an individual link, 
showing that whereas for the case $\sigma=1$ the individual links still show some (noisy) periodicity, for $\sigma=4$ any sign of periodicity has been washed off by the noise. The right panels display the scalar embedding of the full network trajectory using the classical-MDS strategy (embeddings from the PCA-based strategies are similar or marginally poorer). The three viable strategies (both PCA-based embeddings and the one based on classic-MDS) are capable to capture the periodic nature of the network trajectory even when at the individual link dynamics the trajectory is essentially indistinguishable from noise.}
\end{figure}

\noindent For each case, we generate a pulsating network trajectory of $T=500$ snapshots (where each snapshot has $N=20$ nodes), compute the distance matrix on the network trajectory, and perform the battery of scalar embeddings. We summarise results in Fig.~\ref{fig:pulsating}. The middle panels display the time series of an individual link, 
certifying that 
for $\sigma=1$ the individual links still show some (noisy) periodicity and for $\sigma=4$ any sign of periodicity has been washed off by the noise. The right panels 
display the scalar embedding of the full network trajectory (for illustration, we focus on the classic-MDS embedding, --the one providing best results in this case-- although the two PCA-based strategies are also capable of detecting the periodic nature.  
Interestingly, not only in the $\sigma=1$ but also for $\sigma=4$, the scalar embeddings clearly {\it enhance} the periodic backbone of the network dynamics. 
Overall, results suggest that the scalar embeddings induced by both PCA and classic-MDS strategies are capable of inheriting the pulsating character of the temporal network and filtering out {noise}, thus enhancing the signal{-}to{-}noise ratio. 


\subsubsection{{Periodic temporal networks: effects of fixed vs varying number of links, size and edge density, proportion of links with periodic activity and detection of change points}} \label{sec:results_periodic}
{In this section, we investigate the following four questions by additional numerical experiments: (i) does the scalar embedding capture the abovementioned network's periodicity because the edge density of the snapshot varies periodically, and in general does the scalar embedding capture the fluctuations of edge density? (ii) How do parameters such as network size or the edge density affect results? (iii) Is the embedding capable of retrieving the periodic backbone when not all the network but smaller subsets are the ones with an active periodic structure? (iv) Can the embedding efficiently capture change points?}

\medskip \noindent
{First, to assess the effect of the number of links we construct three further types of periodic network trajectories. {\it Type 1} generates periodic temporal networks where each snapshot is an Erd\H{o}s-Reny\'{i} graph with $N$ nodes and a periodically-varying edge probability $p(t)=[1+\cos(2\pi t)/2]$. In this type, the structure of snapshots is uncorrelated, but the average degree of the temporal network $\langle k \rangle = (n-1)p(t)$ varies periodically with period $2\pi$, i.e., all periodicity of this network trajectory can be explained by the (one-dimensional) average degree. {\it Type 2} on the other hand initially builds a subsequence of $P$ stacked Erd\H{o}s-Reny\'{i} graphs generated by $\mathscr{G}(N,p)$. Subsequently, it replicates such exact subsequence of graphs many times to build exact replicas and finally concatenates the replicas, so as to build a long, periodic temporal network with period $P$. In this network trajectory, by construction the average degree of the snapshots within each subsequence is a random variable that fluctuates around its mean value $p(N-1)$, and this fluctuating pattern repeats with period $P$. Finally, {\it Type 3} is similar to {\it Type 2} but uses a slightly different Erd\H{o}s-Reny\'{i} random graph model usually called $\mathscr{G}(N,M)$, so that each of the $P$ generated networks have $N$ nodes and exactly a {\it fixed} number of links $M$ so that the average degree is exactly $\langle k \rangle = 2M$. Therefore, the one-dimensional network metric $\langle k \rangle$ does not explain the periodicity of the trajectory.}\\
{\noindent All three model types above generate periodic network trajectories.
We have generated instances of the three model types  and in every instance we find that both PCA-based and the classical-MDS embeddings correctly recover the periodic nature of the trajectory, even when the number of links is constant (see Fig.~\ref{fig:periodic_revision} in Appendix C). Anecdotally, we find that when the network periodicity can be explained solely by a fluctuating edge density (Type-1), all scalar embeddings perfectly correlate with $\langle k \rangle(t)$ (Pearson correlation $\sim 0.999$), while for Type-2 where fluctuations of edge density have some information but not all, the PCA-based scalar embedding correlates with such time series (Pearson correlation $\sim 0.99$) but the MDS-based embedding does not (Pearson correlation $\sim 0.25$).}

\medskip \noindent {Second, we have analysed whether the network size $N$ (the number of nodes of each snapshot) and the edge density (governed in the models by edge probability $p$ or by the number of links $M$) affect the ability of the embedding to capture the periodicity. Using the {\it Type-3} model of periodic network trajectory, we fix the number of snapshots $T$ and period $P$, and vary $N$ and the edge density $2M/N(N-1)$. We assess whether periodicity can be captured in the scalar embedding by comparing the value of the autocorrelation function at the true period $\tau = P$ with respect to all the values $0<\tau <P$. To this end, we define a $z$-score:}
{
\begin{equation}
z=\frac{\text{ACF}(\tau=P)- \langle \text{ACF}(\tau) \rangle_{\tau < P} }{ \sigma( \text{ACF}(\tau))_{\tau < P} )},
\label{eq:z}
\end{equation}
where $\langle \text{ACF}(\tau) \rangle_{\tau < P} =\frac{\sum_{\tau=1}^{P-1}\text{ACF}(\tau)}{P-1}$ and $\sigma( \text{ACF}(\tau))_{\tau < P} )$ is the standard deviation of all the values of the autocorrelation function for lags larger than 0 and smaller than the true period $P$. We use $z>3$ as a criterion of correct detection of periodicity \cite{Correlations2022Lucas}. Results are depicted in Fig.~\ref{fig:heatmap_periodicity}, and indicate that detection of periodicity is systematically possible and fairly independent of the network size and edge density.}

\medskip \noindent {Third, to assess how detectability is affected when the links contributing to the periodic pattern varies, we develop yet another model. We start from the model generating white network trajectories introduced in Sec.~\ref{sec:white} but using $\mathscr{G}(N,M)$ instead of $\mathscr{G}(N,p)$, in order to fix the number of links in each snapshot. This model generates a total of $T$ i.i.d. snapshots with adjacency matrices with entries $\{A_{ij}(1),A_{ij}(2),\dots,A_{ij}(T)\}$. Then, we fix a period $P$ and select at random a set of entries  $E_{\text{pattern}}=\{(u,v)\}$. Finally, we update these entries in every adjacency matrix to match the corresponding entries of the  `period', that is to say: $A_{uv}(\ell+kP)=A_{uv}(\ell)$, for $\ell=1,2,\dots,P$, $k \in \mathbb{N}^+$ and $(u,v)\in E_{\text{pattern}}$. By doing this, the initially white network trajectory now has a periodic component of period $P$, but only present in a subset of entries of the adjacency matrix. When the cardinality $\text{card}(E_{\text{pattern}})=N(N-1)/2$, then this model reduces to a Type-3 periodic model. When $\text{card}(E_{\text{pattern}}) < N(N-1)/2$, it should be harder to detect periodicity of the network trajectory. We use the percentage $$\rho = \frac{100\cdot \text{card}(E_{\text{pattern}})}{N(N-1)/2}$$ as the control parameter. We quantify the detectability as in the previous case, i.e., by using Eq.~\eqref{eq:z}. Results applied to a network trajectory with snapshots of $N=20$ nodes are plotted in Appendix C Fig.~\ref{fig:z_rho}, certifying that periodicity is detectable even when the periodic activity is present in as few as $10\%$  of the edges.}

\medskip \noindent {Fourth, to assess whether the scalar embeddings can efficiently capture change points, we have built a composite network trajectory with $T=500$ snapshots. Specifically, we have generated the first 250 snapshots by the Type-3 model with $N=50$, $M=200$ and $P=50$ and the subsequent 250 snapshots by the same model but with $P=60$. We remark that all snapshots have the same number of nodes and links. We have carried out the scalar embedding of the network trajectory using the PCA-embedding procedure; results are similar for PCA-projection and classic-MDS procedures. Figure~\ref{fig:change} displays the resulting scalar embedding $z_t$. We find that the scalar signal $z_t$ changes at $t=250$, recovering the change point between the two periodic network models (i.e., from $P=50$ to $P=60$). The autocorrelation function calculated for the first and second half of the scalar signal, shown in  Fig.~\ref{fig:change}B and C, respectively, certifies that each half of the scalar signal is periodic with the adequate period ($P=50$ and $60$, respectively).}

\medskip \noindent As a final note, we have also checked that oscillatory or periodic dynamics composed by more than one period --i.e. multiscale dynamics-- is also correctly captured by the scalar embedding, as its autocorrelation function correctly captures the presence of peaks at the natural periods and combinations of periods (results not shown).

\begin{figure}[htb!]
\centering
\includegraphics[width=0.95\linewidth]{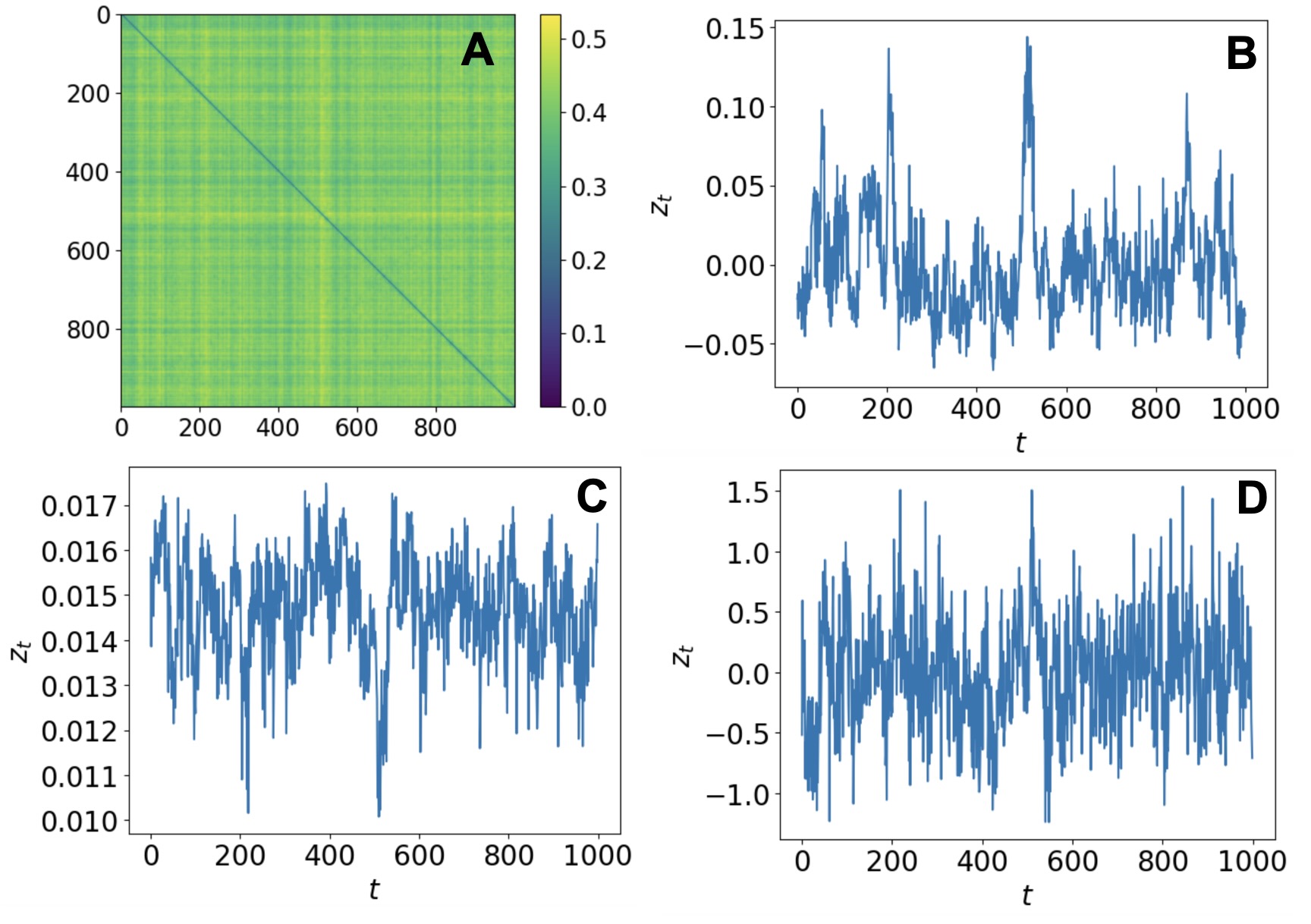}
\caption{\label{fig:DARN3signal} {\bf Scalar embedding of DARN(p) network trajectories. } Scalar embedding of a correlated temporal network trajectory of $T=10^3$ network snapshots generated by a DARN(3) model with parameters $(q,y)=(0.6,0.1)$ and $N=20$ nodes per snapshot. (A) Distance matrix between network snapshots, based on Eq.~\ref{eq:d_snapshots}.
Panels (B)--(D) report the scalar embedding $(z_t)_{t=1}^T, z_t\in \mathbb{R}$ of the network trajectory based on the three viable procedures: (B) {c}lassical-MDS embedding, (C) PCA-embedding, and (D) PCA-projection.} 
\end{figure}

\begin{figure}[htb!]
\centering
\includegraphics[width=0.95\linewidth]{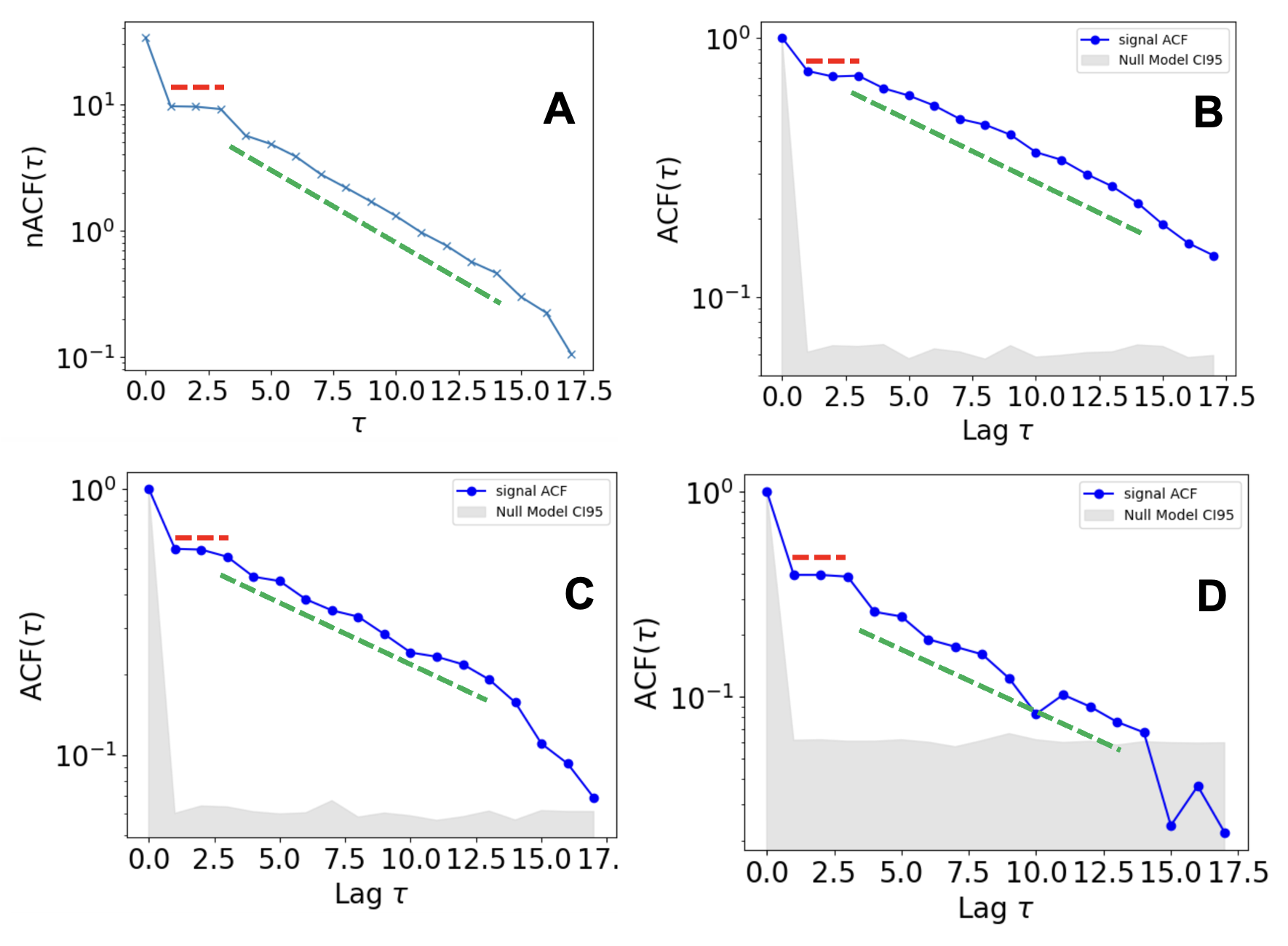}
\caption{\label{fig:DARN3ACF} {\bf Autocorrelation analysis of DARN(p) network trajectories and its embeddings.}  (A) Semi-log plot of the Network Autocorrelation Function $\text{nACF}(\tau)$ estimated from the network trajectory $(G_t)_{t=1}^{1000}$,  where each snapshot is a network of 20 nodes) generated by the DARN(3) model reported in Fig.~\ref{fig:DARN3signal}. 
The network trajectory has memory order $p=3$, and thus we have $\text{nACF}(1)=\text{nACF}(2)=\text{nACF}(3)$, followed by an {exponentially decaying} tail for $\tau>p$, {as indicated by the dashed lines}. {Panels (B)--(D)}  report a semi-log plot of the (scalar) autocorrelation function $\text{ACF}(\tau)$ of the scalar time series $(z_t)_{t=1}^{1000}$ (alongside a gray area reporting the $95\%$ confidence interval of a null model which consists in computing such scalar autocorrelation function in an ensemble of $10^3$ times shuffled signals). Each panel reports results for the different scalar embeddings of Fig.~\ref{fig:DARN3signal}: (B) obtained via  classic-MDS embedding,(C) obtained via PCA-embedding, (D) obtained via PCA-projection.   All scalar embeddings seem qualitatively capture the two components of the autocorrelation function, with a specially good performance by the PCA-embedding and PCA-projection strategies.}
\end{figure}

\begin{figure}[htb!]
\centering
\includegraphics[width=0.95\linewidth]{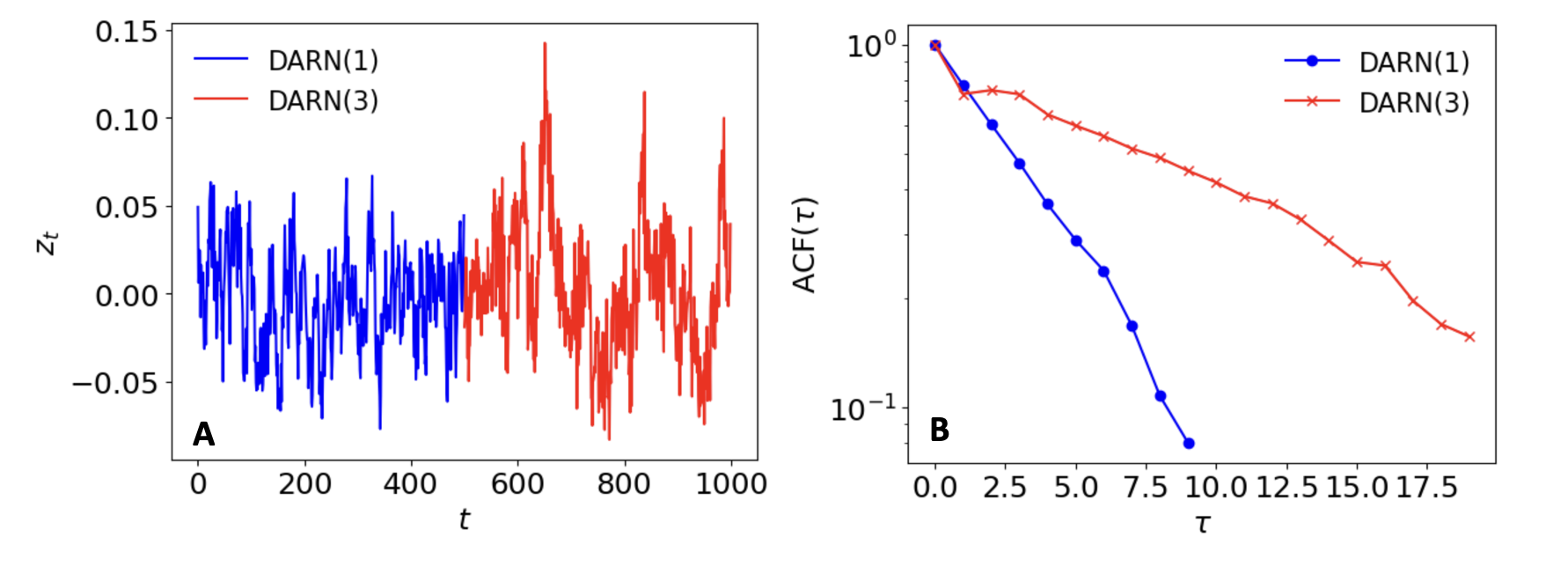}
\caption{\label{fig:change_darn} {{\bf Change point detection in the scalar embedding of mixtures of DARN($p$) network trajectories.}   (A) Scalar embedding $z_t$ of a mixture of two DARN($p$) processes: a DARN($1$) network trajectory of 500 snapshots (blue) and a DARN($3$) network trajectory of 500 snapshots (red). The full network trajectory is built by stacking both collection of snapshots, and the scalar embedding is applied to the full trajectory. One can already see that there is a change point at $t=500$.
(B) Semi-log plot of autocorrelation function $\textbf{ACF}(\tau)$ of the subsignal up to the change point $(z_1,\dots,z_{500})$ (blue) and after the change point $(z_{500},\dots,z_{1000})$ (red), recovering the fingerprint of DARN(1) and DARN(3), respectively. We have used the classic-MDS embedding procedure, but those with the PCA-based embedding are qualitatively the same.}}
\end{figure}

\subsubsection{Network trajectories with memory: DARN(p)}
Moving on, we now consider a model that generates temporal network trajectories with memory:  the Discrete Autorregressive Network model DARN(p) \cite{williams2019effects, williams2022shape, Correlations2022Lucas}. In this model, the dynamics of each link $\ell_t$ is constructed independently, such that with probability $q$, $\ell_{t+1}$ samples uniformly from its past $p$ states, and with probability $(1-q)$, it assigns a Bernoulli trial with probability $y$. 
In other words, when the link update is random, we flip a biased coin and assign the entry $1$ (link present) with probability $y$ and the entry $0$ (link absent) with probability $1-y$).\\ Such process generates a non-Markovian network trajectory with memory order {$p$}. As ground truth, we use the network autocorrelation function $\text{nACF}(\tau)$, which for DARN($p$) processes has a constant value for lags $\tau\le p$ and an exponentially decaying curve for larger lags $\tau >p$, against which we will compare the (scalar) autocorrelation function of the scalar embedding.

\medskip \noindent 
As an {initial} validation, we set $p=3$ and proceed to generate a network trajectory of $T=10^3$ snapshots of a DARN(3) model, where each snapshot has $N=20$ nodes and the model parameters are $(q,y)=(0.6,0.1)$. In Fig.~\ref{fig:DARN3signal}{(A)} we plot a heatmap of the distance matrix ${\cal D}$ constructed from the network trajectory. {Figure~}\ref{fig:DARN3ACF}{(A)} displays the network autocorrelation function of this network trajectory, displaying the abovementioned characteristic features of DARN(p) models.  The different scalar embeddings are reported in panels (B-D) of Fig.~\ref{fig:DARN3signal}, namely scalar embeddings based on: classical-MDS (B), PCA-embedding (C), and PCA-projection (D). Their respective (scalar) autocorrelation functions are reported in panels (B-D) of Fig.~\ref{fig:DARN3ACF}. From these results we can conclude 
that 
the fingerprints of the network autocorrelation function are approximately inherited by all three  types of scalar embedding, with a specially good recovery of both the flat-shape for $\tau \le p$ followed by the exponential decay found in both PCA-based embeddings.

\medskip \noindent {As a second validation, we inspect whether the scalar embedding is capable of correctly inheriting changes in the dynamics of DARN($p$) models, i.e. whether the scalar embedding inherits change points. To test this, we initially construct two network trajectories of $T=500$ snapshots: one generated by a DARN(1) model, and another generated by a DARN(3) model. In both cases, each snapshot has $N=20$ nodes and is generated with the same parameters $(q,y)=(0.6,0.1)$. To model a change point in the network dynamics, we now concatenate the two network trajectories one after the other, yielding a network trajectory with $T=10^3$ snapshots. We compute the scalar embedding of the concatenated network trajectory. The resulting embedding with the classic-MDS procedure is shown in Panel A of Fig.~\ref{fig:change_darn}; results are similar with the PCA-based procedure. The embedding clearly changes its variance around the change point $t=500$. In Panel B of the same figure we plot the autocorrelation function $\text{ACF}(\tau)$ computed from the scalar embedding before (blue) and after (red) the change point. We recover the expected shape for the embedding's autocorrelation of a DARN(1) and DARN(3) models, respectively. Specifically, $\text{ACF}(\tau)$ exponentially decays for $p=1$; it is constant for $\tau \le 3$ followed by a slower exponentially decay for $p=3$. These results confirm that the scalar embedding can adequately inherit not only network trajectories with memory, but mixtures thereof built by using change points.}


\begin{figure}[htb!]
\centering
\includegraphics[width=0.99\linewidth]{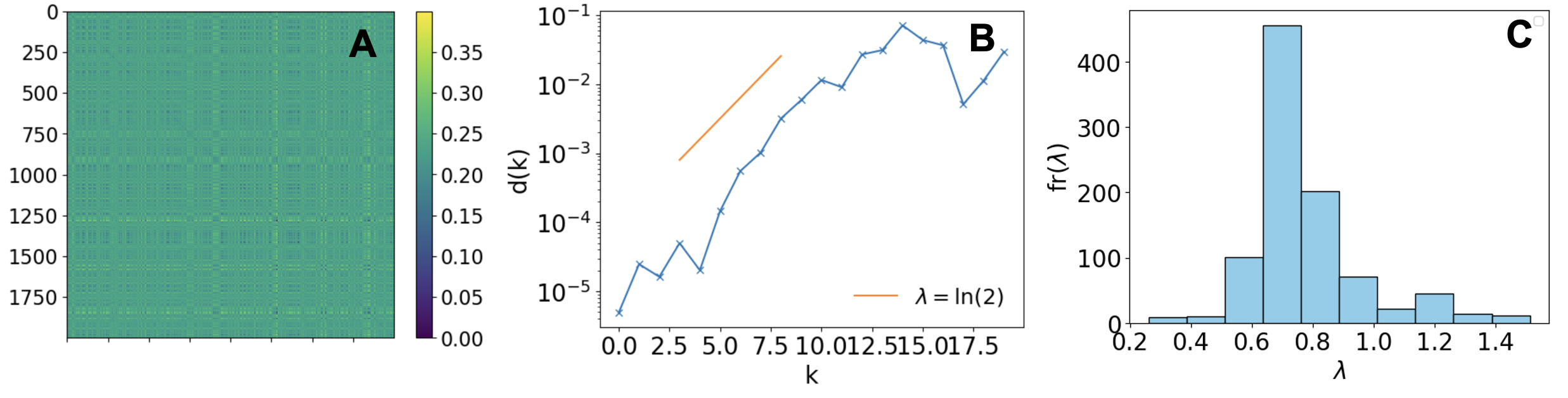}
\caption{\label{fig:chaotic} {\bf Scalar {e}mbedding of a chaotic network trajectory.}  (A) Distance matrix between network snapshots, for a chaotic network trajectory of $T=2000$ snapshots (where each snapshot has $N=500$ nodes) generated via the dictionary
trick, based on a fully chaotic logistic map with theoretical largest Lyapunov exponent $\lambda=\ln 2$. 
(B) Semi-log plot of the distance $d(k)=|z_{u+k}-z_{v+k}|$ between a pair of initially close points $z_u$, $z_v$ in the scalar embedding, with $|z_u-z_v|<10^{-4}$. The distance $d(k)$ increases exponentially (until saturation) with a characteristic exponent close to $\ln 2$. (C) Histogram of characteristic exponents obtained by repeating the procedure in panel (B) for a total of $10^3$ initial conditions, bootstrapped from the scalar embedding. The average of the distribution is close to $\ln 2$.}
\end{figure}

\subsubsection{Chaotic network trajectories: the dictionary trick}
To round off the theoretical validation, here we illustrate the ability of the scalar embeddings to inherit chaotic properties. To build a chaotic network trajectory, we resort to the so-called dictionary trick \cite{Correlations2022Lucas, Annalisa}. This is a method to generate temporal networks with the same dynamical properties of one-dimensional maps. 
We initially consider a one-dimensional 
dynamics generated by the fully chaotic logistic map $x_{t+1}=4x_t(1-x_t)$. This is an interval map $x\in [0,1]$, so the algorithm proceeds by generating a time series $(x_t)_{t=1}^T$, and symbolising the signal after homogeneously partitioning the interval $[0,1]$ into $Q$ equally-sized cells. In parallel, we construct a network dictionary of $Q$ snapshots $\{G[q]\}_{q=1}^Q$ (observe that this is a set of networks, not a temporal network). This dictionary is generated sequentially: starting from an initial (e.g. Erd{\H{o}}s-R\'enyi) network of $N$ nodes $G[1]$, in each step of the process a unique link rewiring is performed. Iterating such process builds $G[2]$, $G[3]$, etc. Such rewiring needs to follow two strict rules: (i) one cannot select a link which had already been inserted from a previous rewiring, and (ii) the new link cannot be inserted in a place which previously had a link that had eventually been rewired. By following these two rules, {the sequence of generated networks is metrical}: any two $G[s]$ and $G[t]$ are precisely $t-s$ rewirings apart, so ${\left\| G[s] - G[t] \right\|}=|t-s|$. Once the dictionary is built, each cell of the interval $[0,1]$ is matched with a network of the dictionary, so that the first cell is assigned $G[1]$, the second cell is assigned $G[2]$, and so on. Finally, each point of the time series $x_t$ is mapped to a network which we label $G_t$, thereby constructing a temporal network $(G_t)_{t=1}^T$ with the same dynamical properties of $(x_t)_{t=1}^T$.

\medskip \noindent We have applied {this} procedure with $Q=10^4$, and constructed a network trajectory of $T=2000$ snapshots, where each snapshot is a network with $N=500$ nodes. In Fig.~\ref{fig:chaotic}\add{(A),} we show the distance matrix of the network trajectory, displaying an apparent uncorrelated structure. 
After computing its scalar embedding $(z_t)_{t=1}^{T}$ (based for illustration in the PCA-embedding strategy), {Fig.~\ref{fig:chaotic}}(B) displays a semi-log plot of the distance $d(k)=|z_{u+k} -  z_{v+k}|${,} where $z_u$ {and} $z_v$ are two points which are close in the embedding space i.e., $|z_u - z_v|<\epsilon=10^{-4}$ --these are akin to recurrences of the trajectory, as in Wolf's method \cite{Annalisa, wolf1985determining}--. The orange solid line depicts an exponential expansion with slope $\lambda = \ln 2$, the Lyapunov exponent of the fully chaotic logistic map. These results suggest that the scalar embedding has inherited the chaotic properties of the network trajectory.
Indeed, we typically observe exponential expansion of nearby trajectories for almost all initial conditions. Note that here we used the 
{PCA-embedding strategy}, but {the} other {three} embedding strategies work reasonably well, {with the caveat that} the metric-MDS one requires antiphase correction.
{Figure~\ref{fig:chaotic}}(C) displays the distribution $P(\lambda)$ obtained when bootstrapping a total of $10^3$ initial conditions from the scalar embedding (as in Wolf's method). The average of the distribution is the theoretical analogue of the network Maximum Lyapunov Exponent \cite{Annalisa} in the scalar embedding, indeed finding it to be close  to $\ln 2$. 

\subsection{Empirical network trajectories}
We finally illustrate the performance of the scalar embedding in two real (empirical) temporal network trajectories. We have concentrated in the scalar embedding obtained via the PCA-embedding strategy, as it showed a consistently good performance in the previous section. 
\subsubsection{Emails}
\begin{figure}[htb!]
\centering
\includegraphics[width=0.85\linewidth]{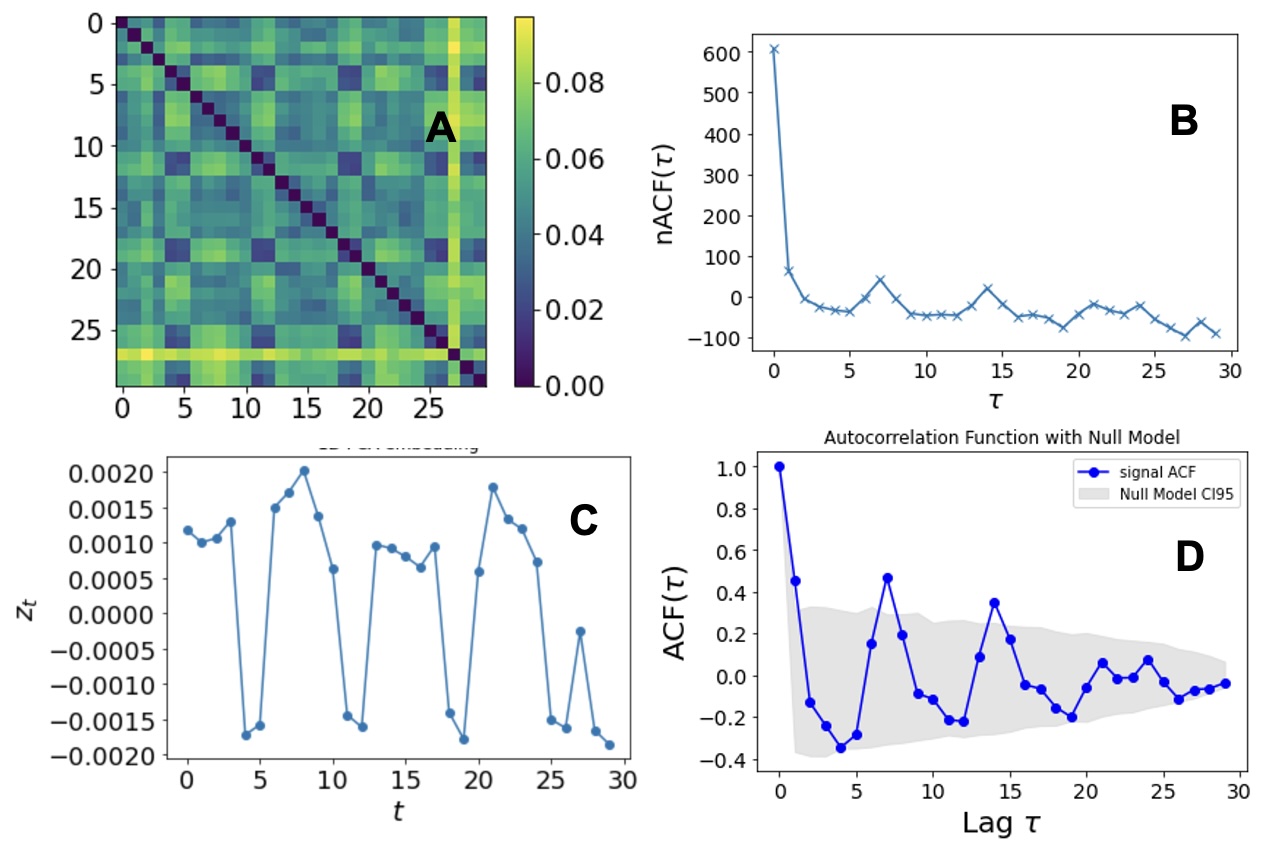}
\caption{\label{fig:emails} {\bf Embedding of the email network trajectory.} 
(A) Distance matrix between network snapshots, for a trajectory of $T=30$ snapshots (one month). Each snapshot aggregates the email activity over a total of one day, between $N=1800$ individuals. $\tau$ scale is in days. (B) Network \add{a}utocorrelation {f}unction $\text{nACF}(\tau)$ estimated from the network trajectory. The function displays a subtle periodic behavior, with period $\tau=7$ days (weekly periodicity).
(C) Scalar embedding of the network trajectory (PCA-embedding). (D) Scalar autocorrelation function $\text{ACF}(\tau)$, where the periodic backbone is clearly enhanced.}
\end{figure}

In Fig.~\ref{fig:emails} we present results on an  empirical email network trajectory\footnote{The 
dataset used \cite{rossi2015network} is called email-dnc \url{https://networkrepository.com/email-dnc.php.}} that was previously identified as having a periodic backbone \cite{Correlations2022Lucas}. 
This is a directed temporal network of emails in the 2016 Democratic National Comittee (DNC). In this network, each node represents a person, and directed edges indicate that one person has sent an email to another. {Each snapshot has $N=1800$ nodes.} Because an email can be sent to multiple recipients, each email is represented by several edges. We have aggregated into the same network snapshot all timestamped edges belonging within a time window of 24 hours. The resulting network trajectory is highly non-stationary, with an initial period of almost no activity. Accordingly, we focus only in a period of the last 30 days (one-month activity, i.e., 30 snapshots), where there was a substantial email exchange. Results in Fig.~\ref{fig:emails} confirm that the scalar embedding of {this} network trajectory captures the periodic backbone. 
{The periodic} fingerprint  appears to be enhanced in {the scalar autocorelation function (see Fig.~\ref{fig:emails}(C))} {with respect to} the nACF case {(Fig.~\ref{fig:emails}(C))} {. This result supports} that the scalar embedding acts as a filter that enhances the signal{-}to{-}noise ratio, {in} agreement with what was observed for synthetic noisy periodic network trajectories.\\
{Now, since the number of links strongly fluctuates over time, we wonder if the scalar embedding is simply capturing a periodicity of the average degree. To remove the effects of this confounding factor --and thus make the task of retrieving dynamics from the scalar embedding substantially more challenging--, we now pollute each snapshot with different amounts of noise (i.e., adding links at random), such that the number of active links is {the same for all the snapshots}. This is akin to superpos{ing} a {\it non-constant} amount of extrinsic noise. The results, depicted in Fig.~\ref{fig:emails_revision} in Appendix C, are qualitatively similar to the ones found in Fig.~\ref{fig:emails}, with the only main difference being that the explained variance of the first principal component drops from $\sim 62\%$ to $\sim 40\%$. In a nutshell, periodicity in this empirical network trajectory is not only given by a periodically fluctuating average degree, and after controlling for this variable, the scalar embedding still captures periodicity. These results align with those obtained for synthetic periodic network trajectories in Sec.~\ref{sec:results_periodic}.}

\subsubsection{Sociopatterns}
\begin{figure}[htb!]
\centering
\includegraphics[width=0.99\linewidth]{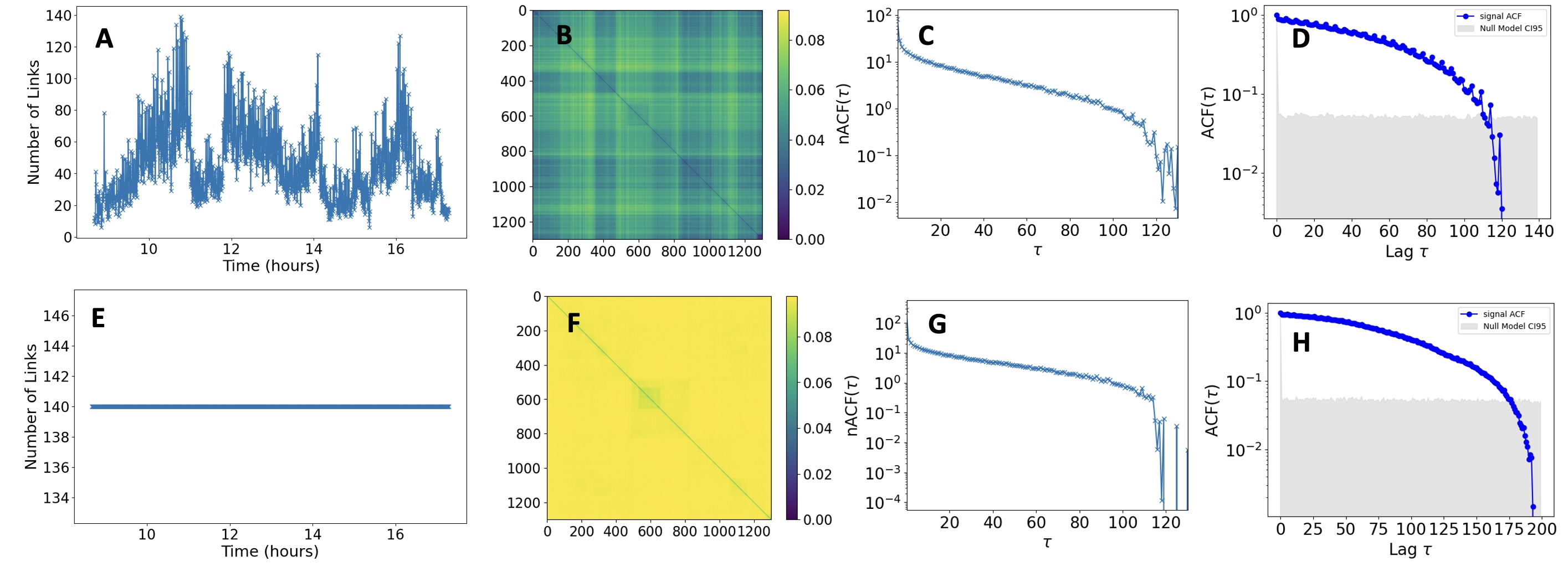}
\caption{\label{fig:primary} {\bf Embedding of primary school network trajectory.}  The network trajectory of $T=1300$ snapshots (roughly equivalent to 8.5 hours of intraday activity in a primary school), where each snapshot has $N=243$ nodes (students) and links model proximity-based interactions among students. This is a subset of the original temporal network, where only one day of activity is selected to remove any source of daily periodicity.
(A) Time series of the number of links of each network snapshot, showing a clear heterogeneous interaction intensity from 9AM to 4PM. (B) The distance matrix of the  network trajectory. (C) Semi-log plot of the network autocorrelation function $\text{nACF}(\tau)$, showing a decaying shape --i.e., memory-- with a characteristic timescale of $\tau=120$ (roughly equivalent to a characteristic time of 47 minutes). 
(D) Semi-log plot of the scalar autocorrelation function $\text{ACF}(\tau)$ of the scalar embedding, showing hints of memory with a characteristic timescale of $\tau=120$ (roughly 47 minutes), equal to the one detected by the network autocorrelation function.
(E) Same as (A), after having polluted the network trajectory with noise of varying intensity, such that each network snapshot has a fixed number of links. 
This removes the source of dynamics based on number of links, and in principle makes the embedding much more challenging.
(F-H) Same as (B-D), for the contaminated network trajectory. 
{Despite} the substantial amount of time-varying noise contamination,  the network autocorrelation function shows the same shape --showing hints of memory-- although with a larger characteristic scale $\tau\approx 175$ (roughly equivalent to 69 minutes).}
\end{figure}

We have further explored the scalar embedding of empirical temporal networks obtained 
from the Sociopatterns project\footnote{\url{www.sociopatterns.org}}. 
In Fig.~\ref{fig:primary} we show the results applied to a primary school temporal network trajectory. This network trajectory has $T=1300$ snapshots (based on the aggregation window, roughly equivalent to 8.5 hours), where each snapshot has $N=243$ nodes 
which correspond to students in a primary school, and links model proximity-based interaction \cite{stehle2011high}. This is a subset of the original temporal network, {and we selected} only one day of {data} to remove any source of daily periodicity. Our results show that the interaction activity strongly fluctuates throughout the day ({Fig.~\ref{fig:primary}(}A{)}) and that the network trajectory displays memory, with a characteristic timescale of about 47 minutes, as found with the network autocorrelation function ({Fig.~\ref{fig:primary}(}C{)}). This characteristic timescale is similar to the typical duration of a lecture.
The scalar embedding based on the PCA-embedding strategy almost perfectly retrieves the dynamics (explained variance of the first principal component is over $92\%$), and its autocorrelation function also displays decaying memory with the same characteristic timescale of 47 minutes ({Fig.~\ref{fig:primary}(}D{)}). 
{We have further performed the link contamination procedure to remove the effect of a varying number of links, see ({Fig.~\ref{fig:primary}(}E{)}). 
We find that the memory structure detected by the network autocorrelation --as well its the characteristic timescale-- is conserved after this drastic contamination ({Fig.~\ref{fig:primary}(}G{)}). This result implies that the memory pattern that we observed cannot be attributed to the periodic fluctuations in the link density.}
As for the scalar embedding, the explained variance of the first principal component decreases but remains substantial (over $17\%$ 
in a system of 1300 principal components). The autocorrelation function still displays decaying memory, although the characteristic timescale is larger.

\section{Discussion}
\label{sec:discussion}
In this work we have explored the performance of four similar strategies --all based on leveraging {well-known} linear dimensionality reduction techniques-- to extract a scalar embedding (i.e., one\add{-}dimensional time series) out of a temporal network trajectory. The common {key} insight underpinning such approaches is assuming that the information which is relevant for capturing the intrinsic dynamics of the temporal network trajectory lies in the relative distance between network snapshots, rather than specific structure within each snapshot, i.e., inter-snapshot information, rather than intra-snapshot one. {This automatically suggests associating to each snapshot a feature vector, where features are the graph distances between that snapshot and every other network snapshot. One can subsequently leverage linear dimensionality reduction techniques --including PCA or MDS-- in slightly different ways, yielding the four embedding procedures explored here.}
We have validated the methods by constructing several synthetic models of network trajectories with different types of dynamics (noisy periodic temporal networks with one or more timescales, temporal networks with memory, chaotic networks, etc), and finding that their scalar embeddings adequately inherit the planted dynamical fingerprints. {We have also provided} some evidence {supporting} that the scalar embedding behaves as a filter that enhances the signal{-}to{-}noise ratio of the original network trajectory. 
The analysis of two empirical network trajectories further confirms the applicability of the method in realistic, high dimensional scenarios.

\medskip \noindent 
Out of the four strategies under analysis, embedding and projection based on PCA and classical-MDS work substantially better than the one based on metric-MDS. There is no clear{ly} superior method among the three successful ones ({i.e.,} PCA-projection, PCA-embedding and classical-MDS), as some work slightly better than others in different cases. However, PCA's explained variance of the first principal component can be used as a quantification of how much information is discarded in the dimensionality reduction, and this might give these strategies an interpretability advantage.
Observe that despite the severe dimensionality reduction, the scalar embedding captures substantial and nontrivial dynamical properties of the original network trajectory. All in all, we thus consider the conceptual approach advanced here to be promising
and offers an avenue for making time series analysis of temporal networks, with high applicability across the disciplines.

\medskip \noindent 
Let us now {reflect on our results, outline some limitations,  open questions for further work and discuss possible applications of the framework}.\\ 
First, note that each snapshot of a temporal network can be trivially reduced to a scalar by extracting from the network a particular topological property, e.g. the edge density, the largest eigenvalue of the adjacency matrix, etc. This approach, however, can be seen as trivial as it only uses information from each snapshot. Conversely, our procedure looks at one network snapshot and leverages information from the distance of such snapshot {\it to every other snapshot} to eventually find such scalar embedding.

\medskip \noindent 
Second, 
this paper proposes a proof of concept, focusing on scalar embeddings. In this sense, the method {can be straightforwardly generalized} to dimensions larger than one {--if and when needed--} along the lines described in Section \ref{sec:methods}. {Intuitively, given a specific dynamics, it is sensible to expect that there is a lower bound for the dimensionality of the embedding, below which the dynamics cannot be recovered and that such a lower bound should somehow depend on the number of network snapshots $T$, the size of each snapshot $N$ and the intrinsic complexity of the dynamics. Our experiments however all find that using a scalar embedding seems enough to capture the main dynamical properties of the network trajectory --let it be a periodic backbone, sensitivity to initial conditions, or memory timescales-- finding this for a variety of values of $T$ and $N$ (i.e., such lower bound appears to saturate to one, for the range of network dynamics analyzed here).  
It is an interesting open problem to mathematically unveil these dependences with as much generality as possible.}

\medskip \noindent 
Third, computational efficiency of the method could also be improved in future work, e.g. by considering online versions of the embedding algorithms. {Likewise, here we used specific normalization and scaling procedures in both PCA and MDS-based approaches. It is unclear whether other data preprocessing --e.g. logarithmic transformations-- could further help stabilise the method.}

\medskip \noindent 
Fourth, as a proof of concept we used an Euclidean norm (Eq.~\ref{eq:d_snapshots}) to assess the distance between two network snapshots, but other choices are also possible \cite{wills2020metrics},
{where different aspects of the graph --including node-based features, edge-based features, or a mix-- could be considered in order to build the distance function. Choosing more sophisticated metrics e.g. graph kernels \cite{vishwanathan2010graph} could also allow to apply the method in temporal networks with varying number of nodes.
Related to this, would two isomorphic graph snapshots map onto the same point in the scalar embedding? The current answer is no if we use a graph distance based on the adjacency matrix such as Eq.~\ref{eq:d_snapshots}. However, by defining a distance based on any graph invariant \cite{caligiuri2024characterising} (i.e., graph properties which are invariant under row/column permutations of the adjacency matrix, such as its spectrum), one could readily extend this scalar embedding to unlabeled temporal networks.}

\medskip \noindent 
Fifth, on relation to the feature matrix, in this work we have {treated as `equally important features' all the relative distances of a snapshot to any other snapshot}. 
This assumption is perhaps too strict, as one can argue that the relative distance between closer snapshots (in time) is a more important feature than the relative distance between snapshots that are very far apart in time. Accordingly, a possible improvement for future work could define some `weight decay' (e.g. a kernel) between snapshots based on how close the snapshots are in time. {This kernel could, in turn,} boost the computational efficiency of the whole methodology. 

\medskip \noindent 
Sixth, it would be interesting to consider other dynamical quantifiers beyond linear correlations or dynamical stability. {In general, having a faithful scalar embedding opens up the possibility of performing time series analysis of temporal networks, including e.g. time series irreversibility \cite{lacasa2012time, arola2023irreversibility}, temporal network reducibility and compression \cite{de2015structural, santoro2020algorithmic, vaudaine2024temporal} or temporal network similarity \cite{dall2024embedding}, to cite some.} 

\medskip \noindent 
Seventh, note that the dimensionality reduction approaches considered here are inherently linear. {In this sense, further work should assess the viability of nonlinear techniques \cite{wang2016structural}.}

\medskip \noindent 
{Finally, applications of this framework are widespread and cover problems involving social, socio-technical, ecological or physical networks. Potential examples include to characterise climate networks and fluid-flow networks in oceanic sciences and fluid mechanics \cite{ser2021lagrangian, dijkstra2019networks, iacobello2019lagrangian}, model the evolution of so-called functional connectivity in the brain \cite{Thompson_et_al} or ecological network changes \cite{trojelsgaard2016ecological, hervias2024structure}, find low-dimensional representation of market dynamics via embedding of temporal financial networks \cite{zhao2018stock}, identify patterns of failures or instabilities in energy grids \cite{martinez2023dynamical}, and provide interpretable descriptions of the training process of deep neural networks \cite{danovski2024dynamical}.}

\subsection*{{Appendix A: Exact scalar embeddings of scalar data using Classical-MDS}}

{Here we give rigorous analytical support to the preliminary validation of complex scalar time series (Section 3.1). In particular, we show that, for a one-dimensional time series \(\{x_t\}_{t=1}^T\), the scalar embedding obtained using the Classical-MDS strategy (introduced in Section 2.2) recovers the original signal up to a constant shift and a reflection of all the entries.}

\medskip \noindent 
{We first define the vector $\bm{v}$, or its transpose 
$\bm{v}^\top = (x_1^2,\, x_2^2,\, \dots,\, x_T^2)$, 
i.e., the vector populated by the squared values of the time series. Using \(\bm{v}\), we can write the squared distance matrix \(D^{(2)}\) as 
\begin{equation}
D^{(2)} = \bm{v}\bm{1}^\top - 2\, \bm{x}\bm{x}^\top + \bm{1}\bm{v}^\top,
\end{equation}
where $\bm{1}$ is a vector of ones. Next, we apply the centering matrix 
\begin{equation}
J = I - \frac{1}{T}\bm{1}\bm{1}^\top,
\end{equation}
which has the property that $J\bm{y} = \bm{y} - \bar{y}\bm{1}$ for any vector $\bm{y}$ (with $\bar{y}$ being the mean of $\bm{y}$). In the classical MDS the centered Gram matrix is defined by Eq.~\eqref{B_matrix}, i.e. 
\begin{equation}
B = -\frac{1}{2} J D^{(2)} J.
\end{equation}
By substituting our expression for $D^{(2)}$ we obtain
\begin{equation}
B = -\frac{1}{2}J\Bigl(\bm{v}\bm{1}^\top - 2\, \bm{x}\bm{x}^\top + \bm{1}\bm{v}^\top\Bigr)J.
\end{equation}
By noting that $J\bm{1} = 0$ we see that, when the centering matrix is applied from either side, the terms $\bm{v}\bm{1}^\top$ and $\bm{1}\bm{v}^\top$ vanish:
\begin{equation}
J\bigl(\bm{v}\bm{1}^\top\bigr)J = \bm{0} \quad \text{and} \quad J\bigl(\bm{1}\bm{v}^\top\bigr)J = \bm{0}.
\end{equation}
Accordingly, we are left with
\begin{equation}
B = J\bm{x}\bm{x}^\top J.
\end{equation}
Since \(J\) is linear and symmetric, $J = J^\top$, we can write
\begin{equation}
J\,\bm{x}\bm{x}^\top\,J = (J\bm{x})(J\bm{x})^\top.
\end{equation}
Using
\begin{equation}
J\bm{x} = \bm{x} - \frac{1}{T}\bm{1}\bm{1}^\top \bm{x} = \bm{x} - \bar{x}\bm{1},
\end{equation}
with \(\bar{x} = \frac{1}{T}\sum_{t=1}^T x_t\), we obtain
\begin{equation}
B = (\bm{x} - \bar{x}\bm{1})(\bm{x} - \bar{x}\bm{1})^\top.
\end{equation}
This last result shows that, when the original time series is one-dimensional, $B$ is a rank-one matrix. It follows that its unique (up to sign) nonzero eigenvector recovers the centered data. In other words, for scalar data (regardless of its temporal complexity), the Classical-MDS embedding, which is directly obtained from the eigen-decomposition of $B$, recovers the original signal, up to a constant mean shift and a sign flip of all the entries in $x$.}

\subsection*{Appendix B: Spontaneous sign flipping in metric-MDS}
\begin{figure}[htb!]
\centering
\includegraphics[width=0.95\linewidth]{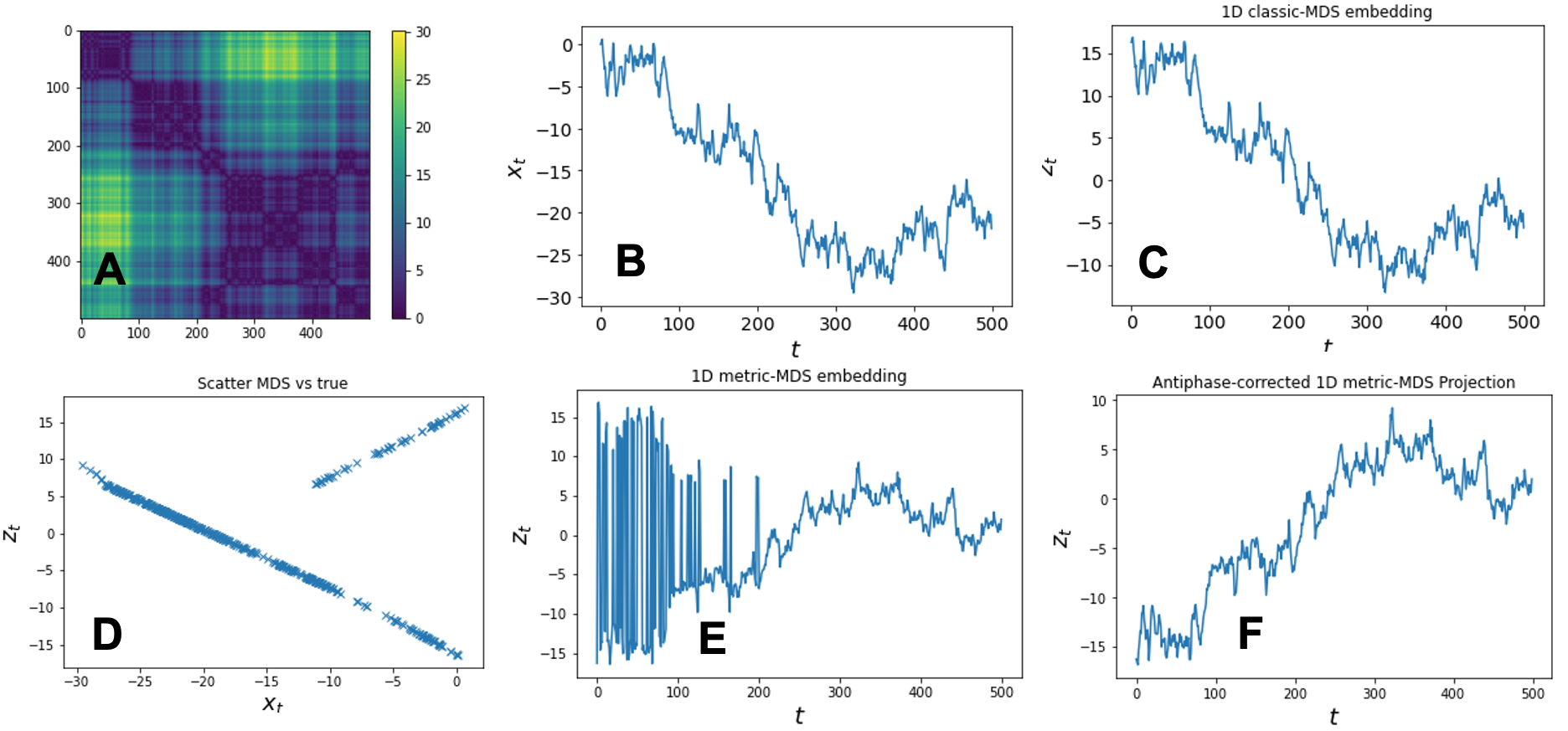}
\caption{\label{fig:antiphase} {\bf Onset and correction of antiphases.}  (A) Distance matrix of a toy one-dimensional random walk trajectory $x_{t+1}=x_t+\xi$. (B) Random walk trajectory $(x_t)_{t=1}^{500}$. (C) Example of scalar embedding $z_t$ that works correctly{. We used the} classic-MDS one, but the {result} is similar with {the} PCA-based ones. 
(D-E) Illustration of the onset of antiphases $z_t\to -z_t$ that emerge for some time values when using metric-MDS based embedding. (D) Scatter plot $z_t$ {vs} $x_t$, where we observe  that several points have a $z_t\to -z_t$ flip.(E) metric-MDS scalar embedding, where we clearly see the position of the antiphases. (F) {A}ntiphase-corrected embedding.}
\end{figure}

The scalar embeddings $z_t$ obtained via metric-MDS sometimes appear to suffer from random, unexpected sign flips $z\to -z$ for some time values $t$. We call these sign flips {\it antiphases}, alluding to the fact that $e^{i\pi}=-1$: sign flip is equivalent to a rotation of $\pi$.
In Fig.~\ref{fig:antiphase} we illustrate this phenomenon in a controlled scenario where we use the metric-MDS strategy to extract the scalar embedding $z_t=\Phi(x_t)$, where $x_t$ is the result of one-dimensional random walk $x_{t+1}=x_t+\xi$. Panel (B) displays the one-dimensional original time series $(x_t)_{t=1}^{500}$. For comparison, Panel (C) displays a correct scalar embedding via classical-MDS. Panel (E) displays the scalar embedding obtained via metric-MDS. Panel (D) depicts a scatter plot between $z_t$ (as obtained via metric-MDS) and $x_t$, detecting the values $x_t$ for which there seems to be a sign flip. These points can affect the subsequent statistical analysis of the embedded trajectory, and thus need to be detected and corrected before any analysis. To do that, we introduce two simple techniques, depending on whether we have access or not to a ground true scalar trajectory $(x_t)$. If we have access to the ground true one-dimensional signal, the antiphase correction is a simple iterative method whereby:
\begin{enumerate}
    \item Initially, we consider a scatter plot $z_t$ vs $x_t$, and we fit a straight line. The quality of the fit is ruined by the points with spontaneous sign flip: correcting these signs will then yield to an improved fit. Therefore:
    \item All the points $\{{x_t}\}$ considered outliers of the linear fit (residual error larger than a certain threshold) are flipped ${x_t}\to - {x_t}$. The changes are accepted if the Pearson's $R^2$ of the scatter plot does not {decrease}.
    \item Step{s} (1) and (2) are repeated until convergence (i.e., no more outliers are detected or the flip does not {increase $R^2$}).
\end{enumerate}
This technique is applied in Fig.~\ref{fig:antiphase}(F). 
Unfortunately, this method cannot be applied {when} we do not have access to $\{x_t\}$ (e.g., for temporal networks). In the latter case, {we can always implement} an antiphase correction that relies on a property of smoothness: if $z_t$ is not in {antiphase} , and if the magnitude of $z_{t+1}$ (i.e., $|z_{t+1}|$) is sufficiently close to $z_{t}$ but its sign is flipped, then it is highly likely that an antiphase took place at $z_{t+1}$, and we perform the flip $z_{t+1}\to -z_{t+1}$. Note, however, that the validity of this method relies on the smoothness assumption. This assumption does not hold e.g. in many  network trajectories.\\ 

\subsection*{{Appendix C: Further analysis}}
\begin{figure}[htb!]
\centering
\includegraphics[width=0.75\linewidth]{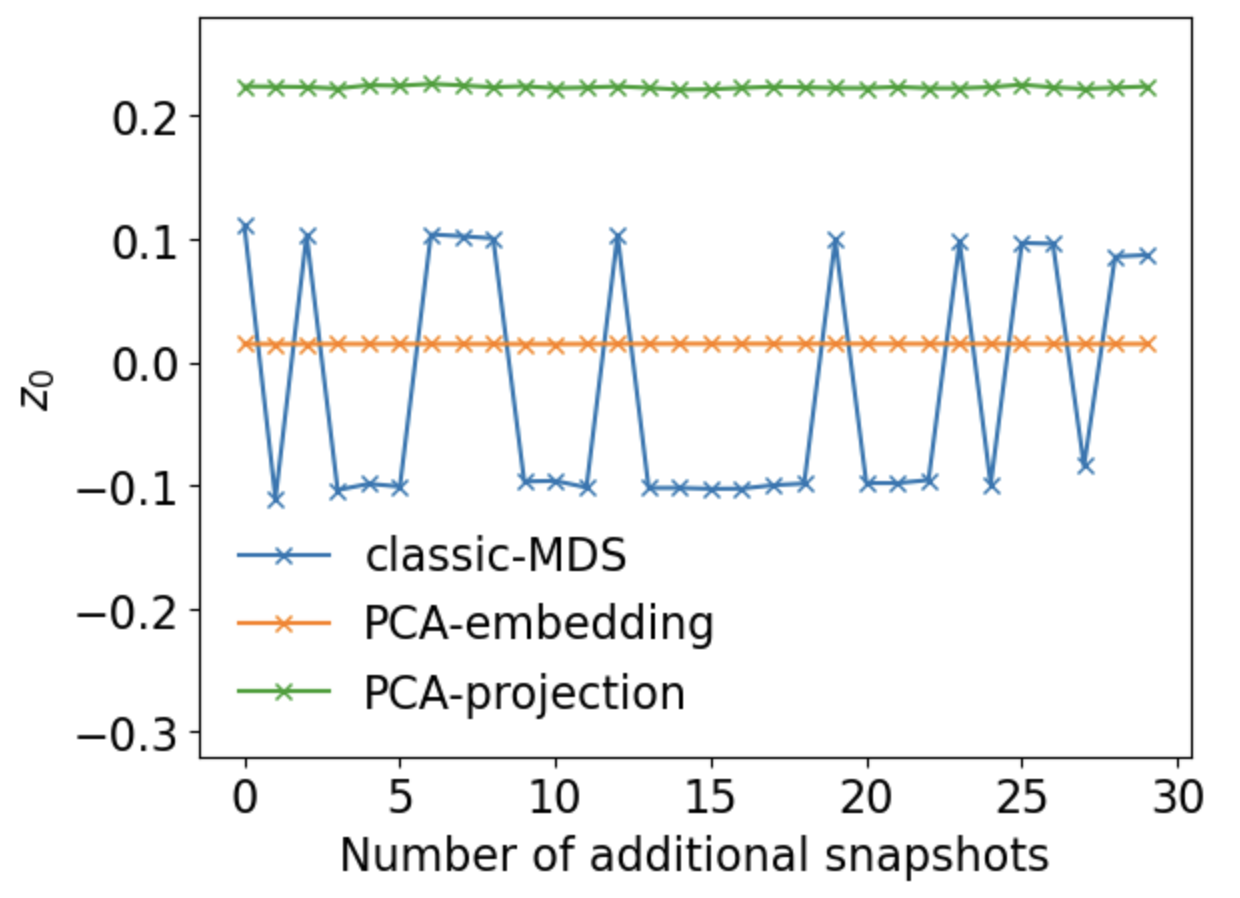}
\caption{\label{fig:stability} {Initial value of the scalar embedding $z_0$ associated to a white network trajectory, as a function of the additional number of snapshots (perturbations), $k$, appended at the end of a network trajectory of $T=500$ snapshots, for the three successful embedding procedures. The PCA-based embeddings are fully stable, whereas for the classic-MDS $z_0$ sometimes flips sign.}}
\end{figure}

\begin{figure}[htb!]
\centering
\includegraphics[width=0.95\linewidth]{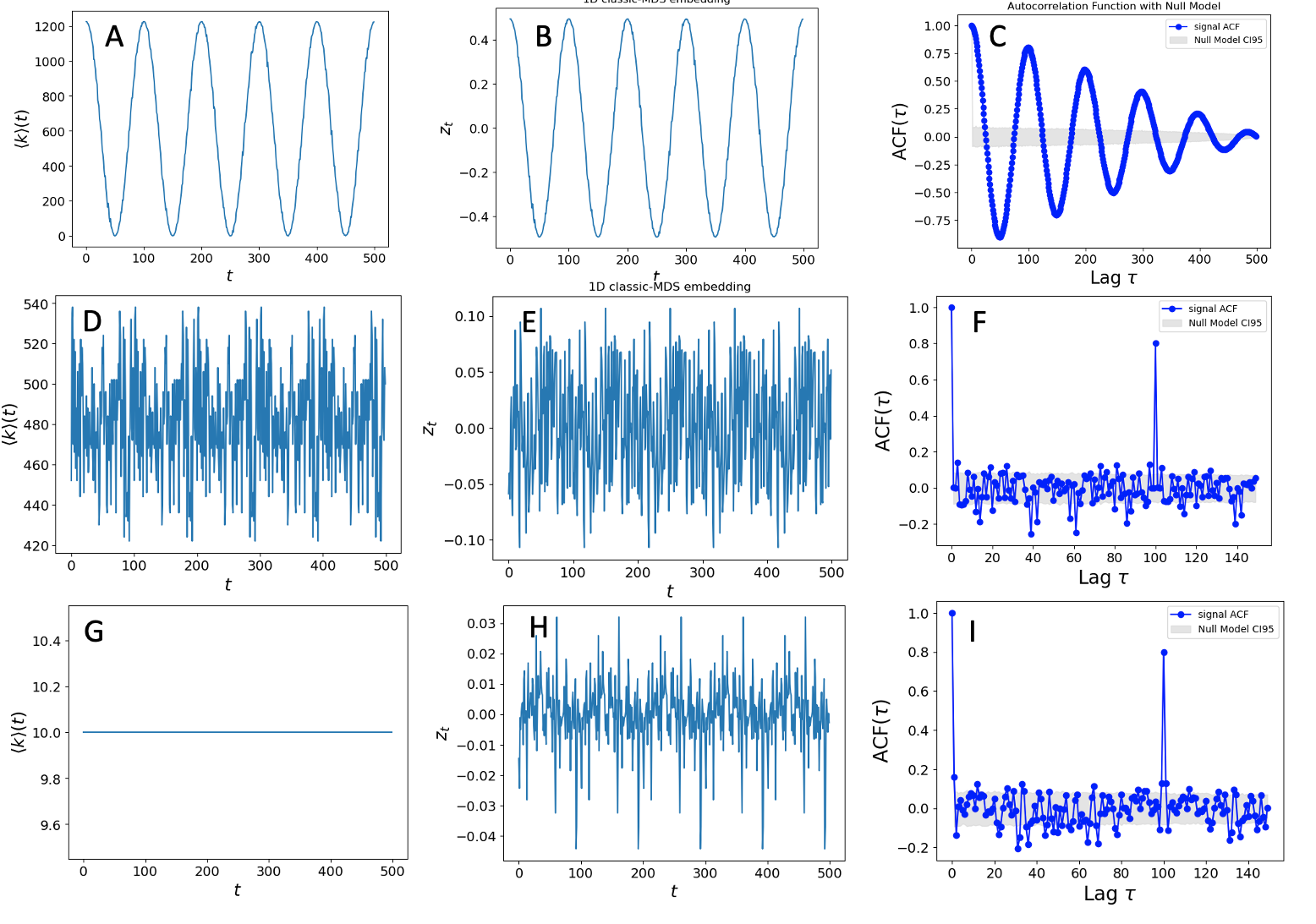}
\caption{\label{fig:periodic_revision} {\bf Scalar embedding captures periodicity also for constant average degree.}  {One-dimensional time series projection of average degree $\langle k\rangle(t)$, CDMS-based scalar embedding $z_t$ and its autocorrelation function for a Type-1 periodic network trajectory (panel A-C), a Type-2 trajectory (panels D-F) and a Type-3 trajectory (panels G-I). All types produce periodic network trajectories with period $100$. Type-1 network trajectories are periodic just because the average degree fluctuates periodically, while in Type-2 and 3 the periodicity is not driven by average degree (in Type-3, $\langle k\rangle(t)$ is constant). In every case the scalar embedding $z_t$ captures the intrinsic periodicity of the network trajectory.}}
\end{figure}

\begin{figure}[htb!]
\centering
\includegraphics[width=0.95\linewidth]{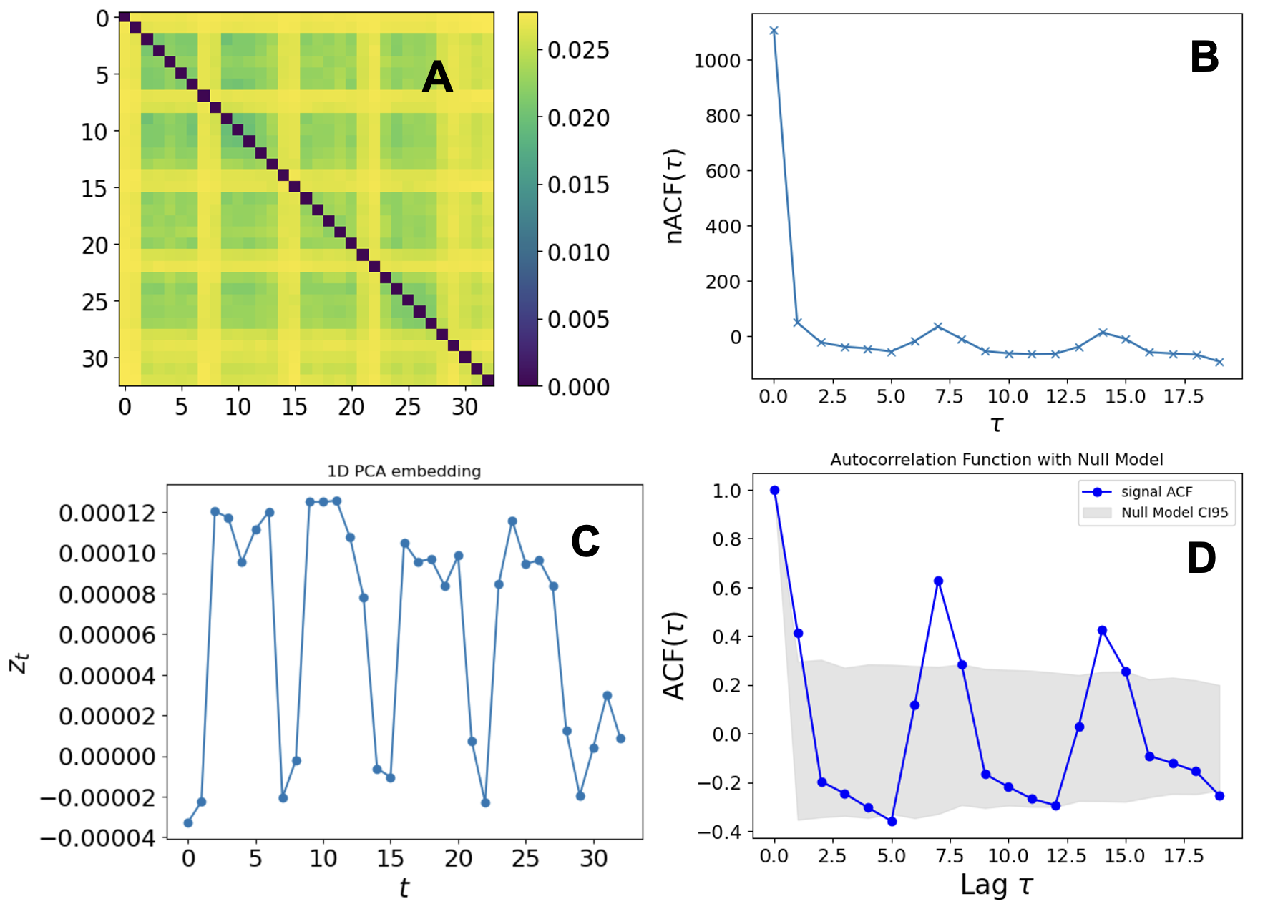}
\caption{\label{fig:emails_revision} {Same as Fig.~\ref{fig:emails}, after polluting the network trajectory with a varying amount of noise such that the total number of links of each snapshot is constant throughout the network trajectory. The distance matrix still manifests some degree of periodicity (which cannot be attributed to a periodically fluctuating number of links), and the scalar embedding still captures the intrinsic periodic pattern.}}
\end{figure}

\begin{figure}[htb!]
\centering
\includegraphics[width=0.65\linewidth]{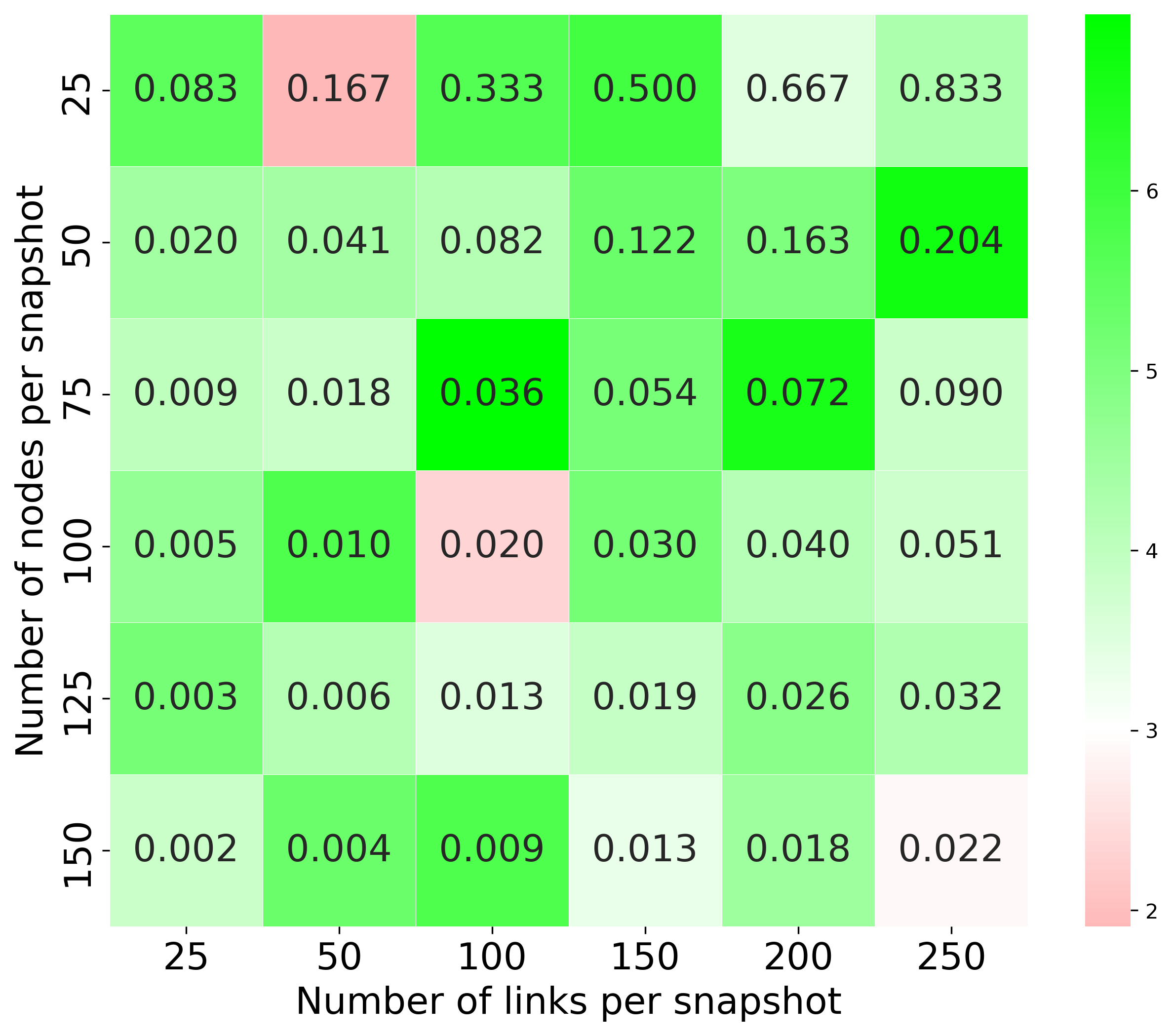}
\caption{\label{fig:heatmap_periodicity} {Detectability of periodic behavior as a function of the number of nodes and the number of links per snapshot, for a Type-3 periodic network trajectory of 100 snapshots with ground true period 20. The detectability is measured in terms of the z-score defined in Eq.~\ref{eq:z}, with a threshold of $z=3$. Results indicate that detectability is not substantially affected by network size or edge density.}}
\end{figure}

\begin{figure}[htb!]
\centering
\includegraphics[width=0.65\linewidth]{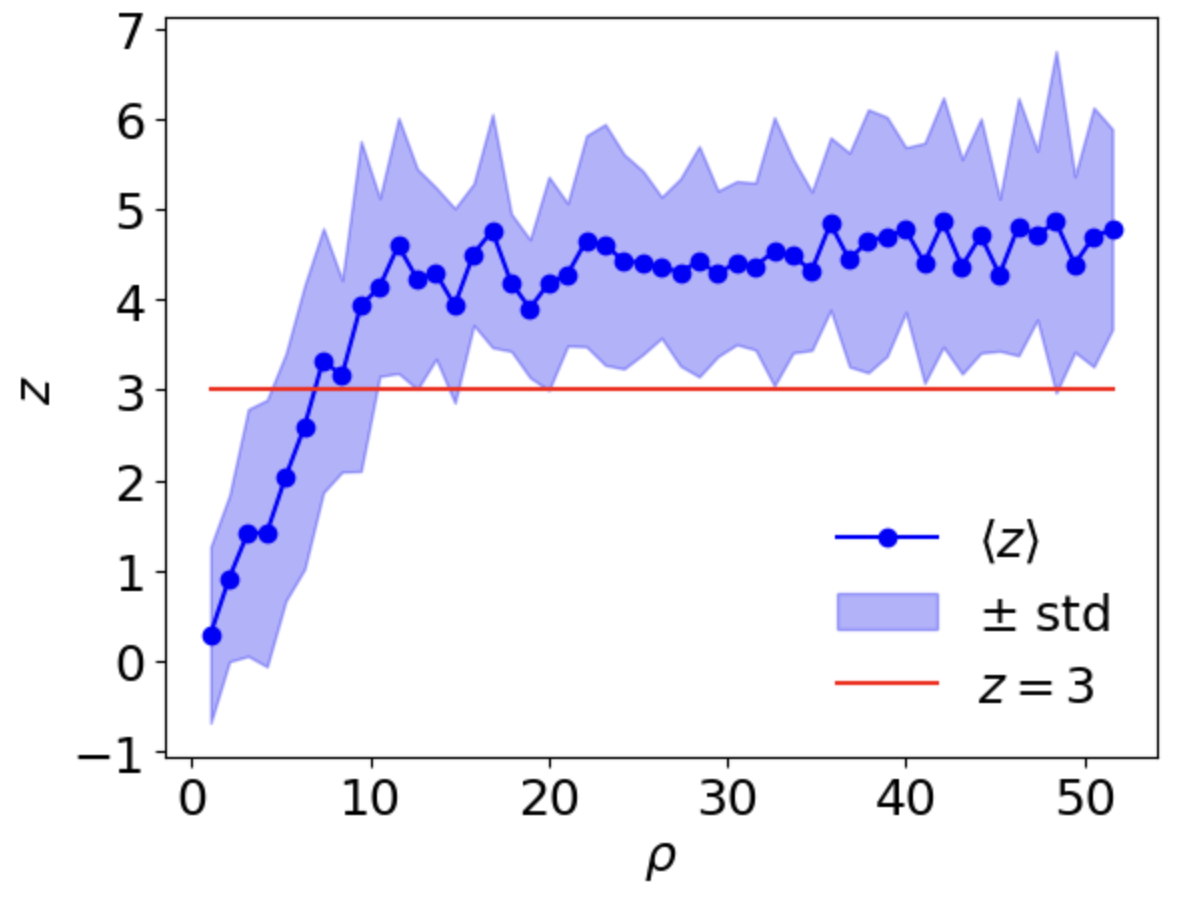}
\caption{\label{fig:z_rho} {Detectability of periodic behavior as a function of the percentage $\rho$ of entries of the adjacency matrix where the periodic activity is concentrated (see Sec.~\ref{sec:results_periodic} for details), for a Type-3 model with $N=20$ nodes per snapshot. The detectability is measured in terms of the z-score defined in Eq.~\ref{eq:z}, with a threshold of $z=3$. Results indicate that detectability is possible already when the periodic activity is concentrated in only $10\%$ of the entries of the adjacency matrices.}}
\end{figure}

\begin{figure}[htb!]
\centering
\includegraphics[width=0.95\linewidth]{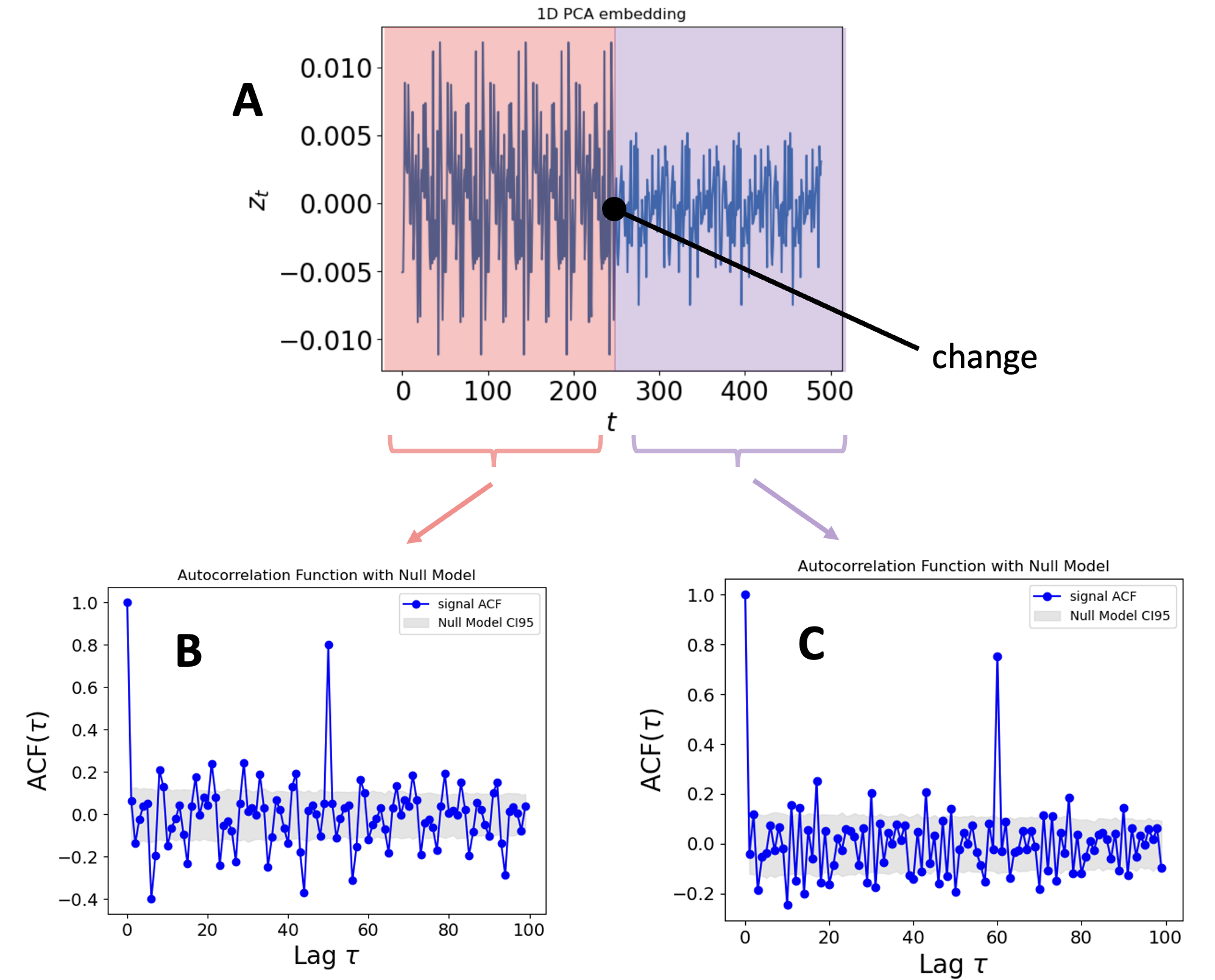}
\caption{\label{fig:change} {(Panel A) PCA-based scalar embedding $z_t$ of a network trajectory ($T=500$ snapshots where each snapshot has $N=50$ nodes are 200 links) generative by two Type-3 models: the first 250 snapshots are generative by a periodic model with period $P=50$, while for  the second 250 snapshots the underlying period is $P=60$. Both models have the same constant number of links. The scalar embedding clearly distinguishes two periodic patterns, with a change point at $t=250$. (Panels B and C) Autocorrelation functions of the first and second parts of the scalar embedding, capturing the periods of the network trajectories in each case.}}
\end{figure}

{This appendix includes Figs.~\ref{fig:stability},~\ref{fig:periodic_revision},~\ref{fig:emails_revision},~\ref{fig:heatmap_periodicity},~\ref{fig:z_rho} and~\ref{fig:change} that report results of additional tests described in the main text.}

\newpage
\clearpage

\noindent {\bf Acknowledgments --} {The authors acknowledge interesting comments from three reviewers.} LL and LAF acknowledge partial support from projects DYNDEEP (EUR2021-122007), MISLAND (PID2020-114324GB-C22) and Maria de Maeztu Seal of Excellence (CEX2021-001164-M) funded by the MICIU/AEI/10.13039/501100011033. {NM acknowledges partial support from the National Science Foundation (grant no. 2052720), the Japan Science and Technology Agency (JST) Moonshot R\&D (grant no. JPMJMS2021), and JSPS KAKENHI (grant nos. JP 23H03414, 24K14840, and 24K030130).}\\

\noindent {\bf Code availability --} All codes will be available upon publication at \url{https://github.com/lucaslacasa}


\end{document}